\documentclass[english,final]{iopart}
\usepackage[T1]{fontenc}
\usepackage[latin9]{inputenc}
\usepackage{geometry}
\usepackage{amstext}

\makeatletter
\usepackage{iopams}
\usepackage{setstack}

\@ifundefined{date}{}{\date{}}
\usepackage{bm}
\usepackage[final]{microtype}
\usepackage{cite}

\makeatother

\usepackage{babel}
\begin{document}

\title{Boundary Algebraic Bethe Ansatz for a nineteen vertex model with
$U_{q}[\mathrm{osp}(2|2)^{(2)}]$ symmetry}

\author{R. S. Vieira}

\ead{rsvieira@df.ufscar.br}

\address{Universidade Federal de São Carlos, Departamento de Física, caixa-postal
676, CEP. 13565-905, São Carlos, SP, Brasil.}

\author{A. Lima Santos}

\ead{dals@df.ufscar.br}

\address{Universidade Federal de São Carlos, Departamento de Física, caixa-postal
676, CEP. 13565-905, São Carlos, SP, Brasil.}
\begin{abstract}
The boundary algebraic Bethe Ansatz for a supersymmetric nineteen
vertex-model constructed from a three-dimensional representation of
the twisted quantum affine Lie superalgebra $U_{q}[\mathrm{osp}(2|2)^{(2)}]$
is presented. The eigenvalues and eigenvectors of Sklyanin's transfer
matrix, with diagonal reflection $K$-matrices, are calculated and
the corresponding Bethe Ansatz equations are obtained.
\end{abstract}

\noindent{\it Keywords\/}: {Algebraic Bethe Ansatz, Open boundary conditions, Lie superalgebras,
twisted Lie algebras, nineteen vertex-models.}
\maketitle

\section{The model}

Over the last decades, great interest has been aroused in the study
of supersymmetric integrable systems. In fact, supersymmetry is now
present in several fields of contemporary mathematics and physics,
ranging from condensed matter physics to superstring theory. We can
cite, for instance, the graded generalizations of Hubbard and \emph{t-J}
models \cite{EsslerKorepinSchoutens1992,EsslerKorepin1992,MartinsRamos1998},
which play an important role in condensed matter physics, and also
the search for solutions of the graded \emph{Yang-Baxter equation}
\cite{Mcguire1964,YangC1967,YangC1968,Baxter1972,Baxter1978,BazhanovShadrikov1987},
which gave origin to important algebraic construction as the supersymmetric
Hopf algebras and quantum groups \cite{Drinfeld1988B}. More recently,
the integrability of supersymmetric models also proved to be important
in superstring theory, more specifically in the AdS/CFT correspondence
\cite{Maldacena1999,MinahanZarembo2003,BeisertStaudacher2003,BenaPolchinskiRoiban2004}. 

The most powerful and beautiful method to analyze these integrable
quantum systems is probably the \emph{Algebraic Bethe Ansatz }\textsc{(aba)}
\cite{TakhtadzhanFaddeev1979,Sklyanin1982,KorepinBogoliubovIzergin1997}.
This technique allows to diagonalize the transfer matrix of a given
integrable quantum system in an analytical way and the commutability
of the transfer matrix (which is guaranteed by the Yang-Baxter equation)
ensures the existence of several conserved quantities in evolution,
the Hamiltonian being one of them. A complete analytical answer for
the problem, however, require the solution of the \emph{Bethe Ansatz
equations}, a system of non-linear equations which have not been completely
solved so far \cite{VieiraLima2015}. The \textsc{aba} was originally
applied to systems with periodic boundary conditions but after the
work of Sklyanin \cite{Sklyanin1988}, integrable models with non-periodic
boundaries could also be handled. Further generalizations \cite{MezincescuNepomechie1991A,BrackenEtAl1998}
showed that the \textsc{aba} can be applied to several classes of
integrable models, described by different Lie algebras and superalgebras,
with both periodic as non-periodic boundary conditions. In the case
of the periodic \textsc{aba}, the fundamental object is a $R$-matrix,
solution of the Yang-Baxter equation, while in the case of the boundary
\textsc{aba}, other fundamental ingredient is necessary: the $K$-matrices
(or \emph{reflection matrices}), which are solutions of the \emph{boundary
Yang-Baxter equations} (\emph{a.k.a.} \emph{reflection equation}s)
\cite{Sklyanin1988,MezincescuNepomechie1991A,BrackenEtAl1998}.

The \textsc{aba} was successfully applied to nineteen vertex models.
In fact, after Tarasov \cite{Tarasov1988} used this technique to
solve the Izergin-Korepin model \cite{IzerginKorepin1981} with periodic
boundary conditions, the Zamolodchikov-Fateev vertex-model \cite{ZamolodchikovFateev1980}
was also solved by Lima-Santos \cite{Lima1999B}. The boundary \textsc{aba}
for these vertex models were performed in \cite{KurakLima2004,KurakLima2005}
together with the supersymmetric $\mathrm{sl}\left(2|1\right)$ and
$\mathrm{osp}\left(2|1\right)$ vertex models. Several other important
results were obtained for nineteen vertex-models $-$ see, for instance,
\cite{VieiraLima2017A} and references therein.

In this work we will study another graded three-state nineteen-vertex
model with reflecting boundary conditions. We derive the boundary
\textsc{aba} for a supersymmetric nineteen-vertex model that was presented
by Yang and Zhang in \cite{YangZhang1999}. The $R$-matrix associated
with this model is constructed from a three-dimensional free boson
representation $V$ of the twisted quantum affine Lie superalgebra
$U_{q}[\mathrm{osp}(2|2)^{(2)}]$ and the periodic \textsc{aba} for
this model was presented in \cite{YangZhen2001}. We would like to
emphasize that vertex-models described by Lie superalgebras $-$ and,
in particular, by twisted Lie superalgebras $-$ are usually the most
complex ones, which of course is due to the high complexity of such
Lie superalgebras \cite{FrappatSciarrinoSorba1989,FrappatSciarrinoSorba2000,KhoroshkinLukierskiTolstoi2001,MackayZhao2001,Ransingh2013,XuZhang2016}.
In fact, even the reflection $K$-matrices of those models were not
yet completely classified, although a great advance in this direction
has been obtained in the last years \cite{Lima2009A,Lima2009B,Lima2009C,Lima2010,VieiraLima2013,VieiraLima2016}.
In the recent work \cite{VieiraLima2017A}, we derived the reflection
$K$-matrices of the Yang-Zhang model which allowed us to implement
now the boundary \textsc{aba} for this model. 

Since we shall deal here with a supersymmetric system, it will be
helpful to remember first the basics of the graded Lie algebras \cite{Kac1977B}.
Let $W=V\oplus U$ be a $\mathbb{Z}_{2}$-graded vector space where
$V$ and $U$ denote their even and odd parts, respectively. In a
$\mathbb{Z}_{2}$-graded vector space we associate a graduation $p(i)$
to each element $\epsilon_{i}$ of a given basis of $V$. In the present
case, we shall consider only a three-dimensional representation of
the twisted quantum affine Lie superalgebra $U_{q}[\mathrm{osp}(2|2)^{(2)}]$
with a basis $E=\{\epsilon_{1},\epsilon_{2},\epsilon_{3}\}$ and the
grading $p(1)=0$, $p(2)=1$ and $p(3)=0$. Multiplication rules in
the graded vector space $W$ differ from the ordinary ones by the
appearance of additional signs. For example, the graded tensor product
of two homogeneous even elements $A\in\mathrm{End}(V)$ and $B\in\mathrm{End}(V)$
turns out to be defined by the formula,
\begin{equation}
A\otimes^{g}B=\sum_{\left\{ i,j,k,l\right\} =1}^{n}\left(-1\right)^{p(i)p(k)+p(j)p(k)}A_{ij}B_{kl}\left(e_{ij}\otimes e_{kl}\right),\label{int.1}
\end{equation}
where $n=3$ is the dimension of the vector space $V$ and $e_{ij}$
are the Weyl matrices ($e_{ij}$ is a matrix in which all elements
are null, except that element on the $[i,j]$ position, which equals
$1$). In the same fashion, the graded permutation operator $P^{g}$
is defined by 
\begin{equation}
P^{g}=\sum_{\left\{ i,j\right\} =1}^{n}\left(-1\right)^{p(i)p(j)}\left(e_{ij}\otimes e_{ji}\right).\label{int.2}
\end{equation}
and the graded transposition $A^{t^{g}}$ of a matrix $A\in\mathrm{End}(V)$
as well as its inverse graded transposition, $A^{\tau^{g}}$, are
defined, respectively, by
\begin{equation}
A^{t^{g}}=\sum_{\left\{ i,j\right\} =1}^{n}\left(-1\right)^{p(i)p(j)+p(i)}A_{ji}e_{ij},\qquad A^{\tau^{g}}=\sum_{\left\{ i,j\right\} =1}^{n}\left(-1\right)^{p(i)p(j)+p(j)}A_{ji}e_{ij},\label{int.3}
\end{equation}
so that $A^{t^{g}\tau^{g}}=A^{\tau^{g}t^{g}}=A$. Finally, the graded
trace of a matrix $A\in\mathrm{End}(V)$ is given by 
\begin{equation}
\mathrm{tr}^{g}(A)=\sum_{i=1}^{n}\left(-1\right)^{p(i)}A_{ii}e_{ii}.
\end{equation}

In the graded case, both the periodic YB equation \cite{Mcguire1964,YangC1967,YangC1968,Baxter1972,Baxter1978,BazhanovShadrikov1987},

\begin{equation}
R_{12}(x)R_{13}(xy)R_{23}(y)=R_{23}(y)R_{13}(xy)R_{12}(x),\label{YBE}
\end{equation}
as well as the boundary YB equation \cite{Sklyanin1988,MezincescuNepomechie1991A,BrackenEtAl1998},
\begin{equation}
R_{12}(x/y)K_{1}^{-}(x)R_{21}(xy)K_{2}^{-}(y)=K_{2}^{-}(y)R_{12}(xy)K_{1}^{-}(x)R_{21}(x/y),\label{BYBE}
\end{equation}
can be written in the same way as in in the non-graded case: it is
only necessary to employ graded operations instead of the usual operations
\cite{BrackenEtAl1998}. 

The $R$-matrix, solution of the graded YB equation (\ref{YBE}),
associated with the Yang-Zhang vertex-model \cite{YangZhang1999}
can be written, up to a normalizing factor and employing a different
notation, as follows: 
\begin{equation}
R(x)=\left(\begin{array}{ccccccccc}
r_{1}(x) & 0 & 0 & 0 & 0 & 0 & 0 & 0 & 0\\
0 & r_{2}(x) & 0 & r_{5}(x) & 0 & 0 & 0 & 0 & 0\\
0 & 0 & r_{3}(x) & 0 & r_{6}(x) & 0 & r_{7}(x) & 0 & 0\\
0 & s_{5}(x) & 0 & r_{2}(x) & 0 & 0 & 0 & 0 & 0\\
0 & 0 & s_{6}(x) & 0 & r_{4}(x) & 0 & r_{6}(x) & 0 & 0\\
0 & 0 & 0 & 0 & 0 & r_{2}(x) & 0 & r_{5}(x) & 0\\
0 & 0 & s_{7}(x) & 0 & s_{6}(x) & 0 & r_{3}(x) & 0 & 0\\
0 & 0 & 0 & 0 & 0 & s_{5}(x) & 0 & r_{2}(x) & 0\\
0 & 0 & 0 & 0 & 0 & 0 & 0 & 0 & r_{1}(x)
\end{array}\right),\label{R}
\end{equation}
where the amplitudes $r_{i}\left(x\right)$ and $s_{i}\left(x\right)$
are given respectively by
\begin{eqnarray}
r_{1}\left(x\right) & = & q^{2}x-1,\\
r_{2}\left(x\right) & = & q\left(x-1\right),\\
r_{3}\left(x\right) & = & q\left(q+x\right)\left(x-1\right)/\left(qx+1\right),\\
r_{4}\left(x\right) & = & q\left(x-1\right)-\left(q+1\right)\left(q^{2}-1\right)x/\left(qx+1\right),\\
r_{5}\left(x\right) & = & q^{2}-1,\\
r_{6}\left(x\right) & = & -q^{1/2}\left(q^{2}-1\right)\left(x-1\right)/\left(qx+1\right),\\
r_{7}\left(x\right) & = & \left(q-1\right)\left(q+1\right)^{2}/\left(qx+1\right),\\
s_{5}\left(x\right) & = & \left(q^{2}-1\right)x=xr_{5}\left(x\right),\\
s_{6}\left(x\right) & = & -q^{1/2}\left(q^{2}-1\right)x\left(x-1\right)/\left(qx+1\right)=xr_{6}\left(x\right),\\
s_{7}\left(x\right) & = & \left(q-1\right)\left(q+1\right)^{2}x^{2}/\left(qx+1\right)=x^{2}r_{7}\left(x\right).
\end{eqnarray}

This $R$-matrix has the following properties or symmetries \cite{VieiraLima2017A}:
\begin{eqnarray}
 & \text{regularity:}\qquad & R_{12}\left(1\right)=f\left(1\right)^{1/2}P_{12}^{g},\label{Sym1}\\
 & \text{unitarity:} & R_{12}\left(x\right)=f\left(x\right)R_{21}^{-1}\left(x^{-1}\right),\label{Sym2}\\
 & \text{super PT:} & R_{12}\left(x\right)=R_{21}^{t_{1}^{g}\tau_{2}^{g}}\left(x\right),\label{Sym3}\\
 & \text{crossing:} & R_{12}\left(x\right)=g\left(x\right)\left[V_{1}R_{12}^{t_{2}^{g}}\left(\eta^{-1}x^{-1}\right)V_{1}^{-1}\right],\label{Sym4}
\end{eqnarray}
where 
\begin{equation}
f\left(x\right)=r_{1}(x)r_{1}\left(\frac{1}{x}\right)=\left(q^{2}x-1\right)\left(\frac{q^{2}}{x}-1\right),\qquad g(x)=-\frac{qx\left(x-1\right)}{\left(qx+1\right)}.
\end{equation}
Here, $t_{1}^{g}$ and $t_{2}^{2}$ mean graded partial transpositions
in the first and second vector spaces, respectively, and $\tau_{1}^{2}$
and $\tau_{2}^{2}$ the corresponding inverse operations. Besides,
$\eta=-q$ is the \emph{crossing parameter} while 
\begin{equation}
M=V^{t^{g}}V=\mathrm{diag}\left(1/q,1,q\right)
\end{equation}
is the \emph{crossing matrix}. 

Solutions of the boundary YB equation (\ref{BYBE}) for this vertex
model were presented recently in \cite{VieiraLima2017A}. The most
general regular diagonal reflection $K$-matrix,
\begin{equation}
K^{-}(x)=\mathrm{diag}\left(k_{1,1}^{-}(x),k_{2,2}^{-}(x),k_{3,3}^{-}(x)\right),\label{K}
\end{equation}
of the Yang-Zhang vertex model $-$ which is of interest in the present
work $-$ has the following entries \cite{VieiraLima2017A}: 
\begin{eqnarray}
k_{1,1}^{-}(x) & = & 1+\frac{1}{2}\left(x^{2}-1\right)\beta_{1,1},\\
k_{2,2}^{-}(x) & = & x\left[\frac{\left(\beta_{2,2}-\beta_{1,1}\right)x+\left(\beta_{1,1}-\beta_{2,2}+2\right)}{\left(\beta_{1,1}-\beta_{2,2}+2\right)x+\left(\beta_{2,2}-\beta_{1,1}\right)}\right]\left[1+\frac{1}{2}\left(x^{2}-1\right)\beta_{1,1}\right],\\
k_{3,3}^{-}(x) & = & x^{2}\left[\frac{\left(\beta_{1,1}-\beta_{2,2}+2\right)-qx\left(\beta_{2,2}-\beta_{1,1}\right)}{\left(\beta_{1,1}-\beta_{2,2}+2\right)x+\left(\beta_{2,2}-\beta_{1,1}\right)}\right]\nonumber \\
 & \times & \left[\frac{\left(\beta_{2,2}-\beta_{1,1}-2\right)-\left(\beta_{2,2}-\beta_{1,1}\right)x}{\left(\beta_{2,2}-\beta_{1,1}-2\right)x+q\left(\beta_{2,2}-\beta_{1,1}\right)}\right]\left[1+\frac{1}{2}\left(x^{2}-1\right)\beta_{1,1}\right],
\end{eqnarray}
where $\beta_{1,1}$ and $\beta_{2,2}$ are the \emph{boundary free-parameters}
of the solutions. Notice that the properties (\ref{Sym1}), (\ref{Sym2}),
(\ref{Sym3}), and (\ref{Sym4}), enjoyed by the $R$-matrix (\ref{R}),
ensure the existence of the\emph{ dual reflection equation}, 
\begin{eqnarray}
 &  & R_{12}\left(\frac{y}{x}\right)K_{1}^{+}\left(x\right)M_{1}^{-1}R_{21}\left(\frac{\eta}{xy}\right)M_{1}K_{2}^{+}\left(y\right)=M_{1}K_{2}^{+}\left(y\right)R_{12}\left(\frac{\eta}{xy}\right)M_{1}^{-1}K_{1}^{+}\left(x\right)R_{12}\left(\frac{y}{x}\right),\nonumber \\
\label{BYBE2}
\end{eqnarray}
as described by Bracken\textit{ }\textit{\emph{et Al.}} \cite{BrackenEtAl1998}.
Besides, the dual reflection matrices $K^{+}(x)$ which are solutions
of the dual reflection equation (\ref{BYBE2}) can determined by the
following isomorphism \cite{BrackenEtAl1998,VieiraLima2017A}
\begin{equation}
K^{+}(x)=K^{-}(\eta^{-1}x^{-1})M,\label{Iso}
\end{equation}
with a new set of boundary free-parameters (say, with $\alpha_{i,j}$
replacing $\beta_{i,j}$). It is to be noticed that special values
for the boundary free-parameters lead to particular reflection matrices,
for instance, the quantum group invariant solutions $K^{-}(x)=I$
and $K^{+}(x)=M$. 

\section{The Boundary Algebraic Bethe Ansatz}

The \textsc{aba} for quantum integrable systems containing diagonal
boundaries was developed by Sklyanin \cite{Sklyanin1988} for integrable
systms described by symmetric $R$-matrices. Menisezcu and Nepomechie
extended Sklyanin's formalism to get account of non-periodic $R$-matrices
\cite{MezincescuNepomechie1991A} and the graded case was developed
further by Bazhanov and Shadrikov \cite{BazhanovShadrikov1987} and
also Bracken et Al. \cite{BrackenEtAl1998}. 

The fundamental ingredient of the boundary \textsc{aba} is the \emph{Sklyanin
transfer matrix},
\begin{equation}
t(x)=\mathrm{tr}_{a}^{g}\left[K^{+}(x)U(x)\right],\label{Trasnfer}
\end{equation}
where $K^{-}(x)$ and $K^{+}(x)$ are any pair of reflection $K$-matrices
satisfying the reflection equations (\ref{BYBE}) and (\ref{BYBE2}),
respectively, and 
\begin{equation}
U(x)=\left[K^{+}(x)T(x)K^{-}(x)T^{-1}(x^{-1})\right],\qquad T(x)=R_{aN}(x)R_{aN-1}(x)\cdots R_{a1}(x),
\end{equation}
are, respectively, the Sklyanin monodromy and the usual periodic monodromy.
Notice that the operators $R_{aq}(x)$ appearing in the expressions
above act in $\mathrm{End}\left(V_{a}\otimes V_{q}\right)$, where
$V_{a}$ denotes the \emph{auxiliary space} and $V_{q}$, for $1\leq q\leq N$,
are in the \emph{quantum spaces} associated to a lattice of $N$ sites
(the graded trace should be taken in the auxiliary space only). In
performing the \textsc{aba}, we consider any pair $\left\{ K^{-}(x),K^{+}(x)\right\} $
of reflection $K$-matrices which do not necessarily need to be related
by the isomorphism (\ref{Iso}).

As commented already in the introduction, he mean feature of the boundary
\textsc{aba} is that the transfer matrix (\ref{Trasnfer}) commutes
with itself for any values of the spectral parameters $x$ and $y$,
that is, 
\begin{equation}
\left[t(x),t(y)\right]=0,\qquad\forall\left\{ x,y\right\} \in\mathbb{C}.
\end{equation}
This means that $t(x)$ can be thought as the generator of infinitely
many conserved quantities in evolution, which justifies the name of
\emph{integrable }to systems that can be solved by the (boundary)
\textsc{aba}. The commutative property of $t(x)$ can be proved from
the unitarity and crossing symmetries of the $R$-matrix plus the
algebra provided by the reflection equations \cite{Sklyanin1988,MezincescuNepomechie1991A,BrackenEtAl1998};
in particular, this implies the integrability of open quantum spin
chains whose Hamiltonian is given by
\begin{equation}
H=\sum_{i=1}^{N-1}H_{i,i+1}+\frac{1}{2}\left.\frac{\mathrm{d}K_{1}^{-}(x)}{\mathrm{d}x}\right\vert _{x=1}+\frac{\mathrm{tr}_{a}^{g}\left[K_{a}^{+}(1)H_{N,a}\right]}{\mathrm{tr}_{a}^{g}\left[K_{a}^{+}(1)\right]},
\end{equation}
where 
\begin{equation}
H_{i,i+1}=\left.\frac{\mathrm{d}}{\mathrm{d}x}\left[P_{i,i+1}R_{i,i+1}(x)\right]\right\vert _{x=1},\qquad1\leq i\leq N-1.
\end{equation}

We remark, that the free boson realization of the $U_{q}[\mathrm{osp}(2|2)^{(2)}]$
quantum Lie superalgebra considered by Yang-Zhang does not have a
classical limit as $q\rightarrow1$ \cite{YangZhang1999}. This particularity
prevent us to study the Gaudin magnets through the off-shell \textsc{aba}
for this model. Nevertheless, other realizations as, for instance,
that one presented in \cite{MackayZhao2001}, could provide other
vertex models with $U_{q}[\mathrm{osp}(2|2)^{(2)}]$ symmetry that
do have a classical limit.

\subsection{The reference state}

In order to find the eigenvalues and eigenvectors of the transfer
matrix through the Bethe Ansatz method, it is necessary to know at
least one of its eigenvectors that is simple enough so that the corresponding
eigenvalue can be directly computed \cite{Sklyanin1988}. This simple
eigenvector is called \emph{reference state}. 

To find the reference state is useful to rewrite the transfer matrix
in the \emph{Lax representation}, that is, as an three-by-three operator
valued matrix, say, 

\begin{equation}
U(x)=\left(\begin{array}{ccc}
A_{1}(x) & B_{1}(x) & B_{2}(x)\\
C_{1}(x) & A_{2}(x) & B_{3}(x)\\
C_{2}(x) & C_{3}(x) & A_{3}(x)
\end{array}\right).\label{U}
\end{equation}
In this representation, the diagonal Sklyanin's transfer matrix (\ref{Trasnfer}),
becomes, 
\begin{equation}
t(x)=k_{1,1}^{+}(x)A_{1}(x)-k_{2,2}^{+}(x)A_{2}(x)+k_{3,3}^{+}(x)A_{3}(x),
\end{equation}
and we can easily verify that the following state 
\begin{equation}
\Psi_{0}=\left(1,0,0\right)^{t}\label{Psi0}
\end{equation}
is an eigenvector of the transfer matrix and, hence, it is the reference
state we were looking for. 

In fact, the action of the monodromy elements over $\Psi_{0}$ can
be evaluated following \cite{KurakLima2004,KurakLima2005}, and reads,
\begin{equation}
A_{i}(x)\Psi_{0}=\alpha_{i}(x)\Psi_{0},\qquad B_{i}(x)\Psi_{0}\not\neq\xi\Psi_{0},\qquad C_{i}(x)\Psi_{0}\not=0\Psi_{0},\qquad1\leq i\leq3.
\end{equation}
where $\xi$ can be any complex number and
\begin{eqnarray}
\alpha_{1}(x) & = & k_{1,1}(x)r_{1}^{2N}(x)\left/f^{N}(x)\right.,\\
\alpha_{2}(x) & = & F_{1}(x)\alpha_{1}(x)+\left[k_{2,2}(x)-F_{1}(x)k_{1,1}(x)\right]r_{2}^{2N}(x)\left/f^{N}(x)\right.,\\
\alpha_{3}(x) & = & \left[F_{2}(x)-F_{1}(x)g_{3}(x)\right]\alpha_{1}(x)+F_{3}(x)\alpha_{2}(x)\nonumber \\
 & + & \left[k_{3,3}(x)-F_{3}(x)k_{2,2}(x)-F_{4}(x)k_{1,1}(x)\right]r_{3}^{2N}(x)\left/f^{N}(x)\right.,
\end{eqnarray}
where,
\begin{eqnarray}
F_{1}(x) & = & s_{5}\left(x^{2}\right)\left/r_{1}\left(x^{2}\right)\right.,\label{F1}\\
F_{2}(x) & = & s_{7}\left(x^{2}\right)\left/r_{1}\left(x^{2}\right)\right.,\label{F2}\\
F_{3}(x) & = & -\frac{r_{1}\left(x^{2}\right)s_{5}\left(x^{2}\right)-r_{5}\left(x^{2}\right)s_{7}\left(x^{2}\right)}{r_{1}\left(x^{2}\right)r_{4}\left(x^{2}\right)+r_{5}(x^{2})s_{5}\left(x^{2}\right)},\label{F3}\\
F_{4}(x) & = & \frac{r_{4}\left(x^{2}\right)s_{7}\left(x^{2}\right)+s_{5}\left(x^{2}\right)s_{5}\left(x^{2}\right),}{r_{1}\left(x^{2}\right)r_{4}\left(x^{2}\right)+r_{5}(x^{2})s_{5}\left(x^{2}\right)}\label{F4}
\end{eqnarray}

Therefore, the action of the transfer matrix on $\Psi_{0}$ reads:
\begin{equation}
t(x)\Psi_{0}=k_{1,1}^{+}(x)\alpha_{1}(x)-k_{2,2}^{+}(x)\alpha_{2}(x)+k_{3,3}^{+}(x)\alpha_{3}(x).
\end{equation}

It will be more convenient, however, to introduce a new set of diagonal
operators, namely,
\begin{eqnarray}
D_{1}(x) & = & A_{1}(x),\\
D_{2}(x) & = & A_{2}(x)-F_{1}(x)D_{1}(x),\\
D_{3}(x) & = & A_{3}(x)-F_{3}(x)A_{2}(x)-\left[F_{2}(x)-F_{1}(x)F_{3}(x)\right]A_{1}(x),
\end{eqnarray}
so that their action on the reference state $\Psi_{0}$ simplifies
to 
\begin{equation}
D_{1}(x)\Psi_{0}=\delta_{1}(x)\Psi_{0},\qquad D_{2}(x)\Psi_{0}=\delta_{2}(x)\Psi_{0},\qquad D_{3}(x)\Psi_{0}=\delta_{3}(x)\Psi_{0},
\end{equation}
 with 
\begin{eqnarray}
\delta_{1}(x) & = & k_{1,1}^{-}(x)r_{1}^{2N}(x)\left/f^{N}(x)\right.,\\
\delta_{2}(x) & = & \left[k_{2,2}^{-}(x)-F_{1}(x)k_{1,1}^{-}(x)\right]r_{2}^{2N}(x)\left/f^{N}(x)\right.,\\
\delta_{3}(x) & = & \left[k_{3,3}^{-}(x)-F_{3}(x)k_{2,2}^{-}(x)-F_{4}(x)k_{1,1}^{-}(x)\right]r_{3}^{2N}(x)\left/f^{N}(x)\right..
\end{eqnarray}
Similarly, the transfer matrix can be rewritten as
\begin{equation}
t(x)=\Omega_{1,1}(x)D_{1}(x)+\Omega_{2,2}(x)D_{2}(x)+\Omega_{3,3}(x)D_{3}(x),
\end{equation}
with 
\begin{eqnarray}
\Omega_{1,1}(x) & = & k_{1,1}^{+}(x)-F_{1}(x)k_{2,2}^{+}(x)+F_{2}(x)k_{3,3}^{+}(x),\label{baba.21-1}\\
\Omega_{2,2}(x) & = & -k_{2,2}^{+}(x)+F_{3}(x)k_{33}^{+}(x),\\
\Omega_{3,3}(x) & = & k_{3,3}^{+}(x),
\end{eqnarray}
 so that their action on the reference state reads now,
\begin{equation}
t(x)\Psi_{0}=\Omega_{1,1}^{+}(x)\delta_{1}(x)+\Omega_{2,2}^{+}(x)\delta_{2}(x)+\Omega_{3,3}^{+}(x)\delta_{3}(x).\label{T}
\end{equation}

\subsection{The 1-particle state}

Once the action of the transfer matrix over the reference state is
determined, we can ask about their excited states. In the framework
of the boundary \textsc{aba}, these excited states are constructed
from the action of the creators operators $B_{1}$ and $B_{2}$ (it
can be verified, however, that the excited states can be constructed
without use the operator $B_{3}$ \cite{Tarasov1988,KurakLima2004,KurakLima2005})
over the reference state $\Psi_{0}$ . Further, we can verify that
the action of $B_{2}$ on $\Psi_{0}$ is proportional to a double
action of $B_{1}$ on $\Psi_{0}$ (in the sense that $B_{2}$ rises
the magnon number of the system twice when compared to $B_{1}$).
From this follows that the first excited state, named here the 1-particle
state, should be defined as,
\begin{equation}
\Psi_{1}\left(x_{1}\right)=B_{1}\left(x_{1}\right)\Psi_{0}.\label{Psi1}
\end{equation}
The new variable $x_{1}$ \textendash{} called \emph{rapidity} \textendash{}
must be determined in order to $\Psi_{1}\left(x_{1}\right)$ be indeed
an eigenvector of the transfer matrix; we shall see below that this
requirement is provided by the Bethe Ansatz equation of the 1-particle
state. 

From (\ref{T}) and (\ref{Psi1}) we can write down the action of
$t(x)$ over $\Psi_{1}\left(x_{1}\right)$: 
\begin{eqnarray}
t(x)\Psi_{1}\left(x_{1}\right) & = & \Omega_{1,1}(x)D_{1}(x)B_{1}\left(x_{1}\right)\Psi_{0}+\Omega_{1,1}(x)D_{2}(x)B_{1}\left(x_{1}\right)\Psi_{0}\nonumber \\
 & + & \Omega_{1,1}(x)D_{3}(x)B_{1}\left(x_{1}\right)\Psi_{0}.
\end{eqnarray}

Notice that we shall need to know the commutation relations between
the diagonal operators $D_{1}(x)$, $D_{2}(x)$ and $D_{3}(x)$ with
$B_{1}\left(x_{1}\right)$ in order to evaluate the action of $t(x)$
on $\Psi_{1}\left(x_{1}\right)$. These commutation relations are
provided by the\emph{ fundamental relation} of the boundary \textsc{aba}
(\ref{FR}) and they are presented in the Appendix. Making use of
the commutation relations (\ref{D1B1}), (\ref{D2B1}) and (\ref{D3B1})
and simplifying the results, we can realize that the action of $t(x)$
on $\Psi_{1}\left(x_{1}\right)$ can be written as follows: 
\begin{equation}
t(x)\Psi_{1}\left(x_{1}\right)=\tau_{1}\left(x|x_{1}\right)\Psi_{1}\left(x_{1}\right)+\beta_{1}\left(x_{1}\right)B_{1}\left(x\right)\Psi_{0}+\beta_{3}\left(x_{1}\right)B_{3}(x)\Psi_{0},\label{Action1}
\end{equation}
where, 
\begin{equation}
\tau_{1}\left(x|x_{1}\right)=a_{1}^{1}(x,x_{1})\Omega_{1,1}(x)\delta_{1}(x)+a_{1}^{2}(x,x_{1})\Omega_{2,2}(x)\delta_{2}(x)+a_{1}^{3}(x,x_{1})\Omega_{3,3}(x)\delta_{3}(x),\label{Eigen1}
\end{equation}
 and,
\begin{eqnarray}
\beta_{1}\left(x_{1}\right) & = & \Omega_{1,1}(x)\left[a_{2}^{1}(x,x_{1})\delta_{1}(x_{1})+a_{3}^{1}(x,x_{1})\delta_{2}(x_{1})\right]\nonumber \\
 & + & \Omega_{2,2}(x)\left[a_{2}^{2}(x,x_{1})\delta_{1}(x_{1})+a_{3}^{2}(x,x_{1})\delta_{2}(x_{1})\right]\nonumber \\
 & + & \Omega_{3,3}(x)\left[a_{2}^{3}(x,x_{1})\delta_{1}(x_{1})+a_{3}^{3}(x,x_{1})\delta_{2}(x_{1})\right],
\end{eqnarray}
\begin{eqnarray}
\beta_{3}\left(x_{1}\right) & = & \Omega_{2,2}(x)\left[a_{4}^{2}(x,x_{1})\delta_{1}(x_{1})+a_{5}^{2}(x,x_{1})\delta_{2}(x_{1})\right]\nonumber \\
 & + & \Omega_{3,3}(x)\left[a_{4}^{3}(x,x_{1})\delta_{1}(x_{1})+a_{5}^{3}(x,x_{1})\delta_{2}(x_{1})\right].
\end{eqnarray}

Now, if $\Psi_{1}(x_{1})$ is an eigenstate of $t(x)$, then we must
have $\beta_{1}^{1}\left(x_{1}\right)=\beta_{1}^{2}\left(x_{1}\right)=0$.
This provides the Bethe Ansatz equation of the 1-particle state that
implicitly fix the rapidity $x_{1}$: 
\begin{eqnarray}
\frac{\delta_{2}(x_{1})}{\delta_{1}(x_{1})} & = & -\frac{\Omega_{1,1}(x)a_{2}^{1}(x,x_{1})+\Omega_{2,2}(x)a_{2}^{2}(x,x_{1})+\Omega_{3,3}(x)a_{2}^{3}(x,x_{1})}{\Omega_{1,1}(x)a_{3}^{1}(x,x_{1})+\Omega_{2,2}(x)a_{3}^{2}(x,x_{1})+\Omega_{3,3}(x)a_{3}^{3}(x,x_{1})}\nonumber \\
 & = & -\frac{\Omega_{2,2}(x)a_{4}^{2}(x,x_{1})+\Omega_{3,3}(x)a_{4}^{3}(x,x_{1})}{\Omega_{2,2}(x)a_{5}^{2}(x,x_{1})+\Omega_{3,3}(x)a_{5}^{3}(x,x_{1})}.\label{BAE1}
\end{eqnarray}
After simplify we can verify that both $\beta_{1}^{1}\left(x_{1}\right)$
as $\beta_{1}^{2}\left(x_{1}\right)$ vanish and also that the right-hand-side
of (\ref{BAE1}) does not actually depend on the spectral parameter
$x$, as it should. The conclusion in that $\Psi_{1}\left(x_{1}\right)$
is an eigenvector of the transfer matrix with eigenvalue $\tau_{1}\left(x|x_{1}\right)$
given by (\ref{Eigen1}).

\subsection{The 2-Particle state}

In the construction of the next excited state $-$ the 2-particle
state $-$ both the operators $B_{1}$ and $B_{2}$ should be taken
into account. This is necessary because both $B_{2}\Psi_{0}$ as $B_{1}B_{1}\Psi_{0}$
are in the same sector (\emph{i.e.,} both states have the same magnon
number). Therefore, the most general 2-particle state should be constructed
through a linear combination of these operators and we can verify
\emph{a posteriori} that the adequate linear combination is as follows:
\begin{equation}
\Psi_{2}\left(x_{1},x_{2}\right)=B_{1}\left(x_{1}\right)B_{1}\left(x_{2}\right)\Psi_{0}+\lambda\left(x_{1},x_{2}\right)B_{2}\left(x_{1}\right)\Psi_{0}=B_{1}\left(x_{1}\right)\Psi_{1}(x_{2})+B_{2}(x_{1})\Psi_{0}.\label{Psi2}
\end{equation}

The coefficient $\lambda\left(x_{1},x_{2}\right)$ of this linear
combination can be fixed by the condition that
\begin{equation}
\Psi_{2}\left(x_{2},x_{1}\right)=B_{1}\left(x_{2}\right)B_{1}\left(x_{1}\right)\Psi_{0}+\lambda\left(x_{2},x_{1}\right)B_{2}\left(x_{2}\right)\Psi_{0}=\omega\left(x_{1},x_{2}\right)\Psi_{2}\left(x_{1},x_{2}\right),
\end{equation}
for some phase function $\omega\left(x_{1},x_{2}\right)$ \cite{Tarasov1988,KurakLima2004,KurakLima2005}.
Making use of the commutation relation (\ref{B1B1}) between $B_{1}(x_{1})$
and $B_{1}(x_{2})$, we get that 
\begin{eqnarray}
\Psi_{2}\left(x_{2},x_{1}\right) & = & b_{1}^{1}\left(x_{2},x_{1}\right)B_{1}\left(x_{1}\right)B_{1}\left(x_{2}\right)\Psi_{0}\nonumber \\
 & + & \left[b_{2}^{1}\left(x_{2},x_{1}\right)\delta_{1}\left(x_{2}\right)+b_{3}^{1}\left(x_{2},x_{1}\right)\delta_{2}\left(x_{2}\right)\right]B_{2}\left(x_{1}\right)\Psi_{0}\nonumber \\
 & + & \left[b_{4}^{1}\left(x_{2},x_{1}\right)\delta_{1}\left(x_{1}\right)+b_{5}^{1}\left(x_{2},x_{1}\right)\delta_{2}\left(x_{1}\right)+\lambda\left(x_{2},x_{1}\right)B_{2}\left(x_{2}\right)\right]\Psi_{0}
\end{eqnarray}
 from which follows that 
\begin{equation}
\omega\left(x_{1},x_{2}\right)=b_{1}^{1}(x_{2},x_{1}),\qquad\lambda\left(x_{1},x_{2}\right)=-b_{4}^{1}\left(x_{1},x_{2}\right)\delta_{1}\left(x_{2}\right)-b_{5}^{1}\left(x_{1},x_{2}\right)\delta_{2}\left(x_{2}\right),
\end{equation}
where we made use of the following properties: 
\begin{equation}
\omega\left(x_{1},x_{2}\right)\omega\left(x_{2},x_{1}\right)=1,\qquad\frac{b_{2}^{1}\left(x_{2},x_{1}\right)}{b_{4}^{1}\left(x_{2},x_{1}\right)}=\frac{b_{3}^{1}\left(x_{2},x_{1}\right)}{b_{5}^{1}\left(x_{2},x_{1}\right)}=-\omega\left(x_{1},x_{2}\right).
\end{equation}

Once $\lambda\left(x_{1},x_{2}\right)$ is determined, we can find
the action of the transfer matrix on the 2-particle state. To this
end, will be necessary to use several other commutation relations
besides the previous one, namely, the commutation relations provided
by (\ref{D1B2}), (\ref{D2B2}), (\ref{D3B2}), (\ref{C1B1}) and
(\ref{C3B1}). Although this computation maybe somewhat extensive,
we can verify that the action of $t(x)$ over $\Psi_{2}\left(x_{1},x_{2}\right)$
can be written as,
\begin{eqnarray}
t(x)\Psi_{2}\left(x_{1},x_{2}\right) & = & \tau_{2}\left(x|x_{1},x_{2}\right)\Psi_{2}\left(x_{1},x_{2}\right)+\beta_{1}^{1}\left(x_{1},x_{2}\right)B_{1}(x)\Psi_{1}\left(x_{2}\right)\nonumber \\
 & + & \beta_{2}^{1}\left(x_{1},x_{2}\right)B_{1}(x)\Psi_{1}\left(x_{1}\right)+\beta_{1}^{3}\left(x_{1},x_{2}\right)B_{3}(x)\Psi_{1}\left(x_{2}\right)\nonumber \\
 & + & \beta_{2}^{3}\left(x_{1},x_{2}\right)B_{3}(x)\Psi_{1}\left(x_{1}\right)+\beta_{12}^{2}\left(x_{1},x_{2}\right)B_{2}(x)\Psi_{0},\label{Action2}
\end{eqnarray}
where,
\begin{equation}
\tau_{2}\left(x|x_{1},x_{2}\right)=\sum_{j=1}^{3}\Omega_{j,j}(x)\delta_{j}(x)a_{1}^{j}(x,x_{1})a_{1}^{j}(x,x_{2}),\label{Eigen2}
\end{equation}
 and
\begin{eqnarray}
\beta_{1}^{1}\left(x_{1},x_{2}\right) & = & \delta_{1}\left(x_{1}\right)a_{1}^{1}\left(x_{1},x_{2}\right)\sum_{j=1}^{3}\Omega_{j,j}\left(x\right)a_{2}^{j}\left(x,x_{1}\right)\nonumber \\
 & + & \delta_{2}\left(x_{1}\right)a_{1}^{2}\left(x_{1},x_{2}\right)\sum_{j=1}^{3}\Omega_{j,j}\left(x\right)a_{3}^{j}\left(x,x_{1}\right),\\
\beta_{2}^{1}\left(x_{1},x_{2}\right) & = & \omega\left(x_{2},x_{1}\right)\beta_{1}^{1}\left(x_{2},x_{1}\right),\\
\beta_{1}^{3}\left(x_{1},x_{2}\right) & = & \delta_{1}\left(x_{1}\right)a_{1}^{1}\left(x_{1},x_{2}\right)\sum_{j=2}^{3}\Omega_{j,j}\left(x\right)a_{4}^{j}\left(x,x_{1}\right)\nonumber \\
 & + & \delta_{2}\left(x_{1}\right)a_{1}^{2}\left(x_{1},x_{2}\right)\sum_{j=2}^{3}\Omega_{j,j}\left(x\right)a_{5}^{j}\left(x,x_{1}\right),\\
\beta_{2}^{3}\left(x_{1},x_{2}\right) & = & \omega\left(x_{2},x_{1}\right)\beta_{1}^{3}\left(x_{2},x_{1}\right),\\
\beta_{12}^{2}\left(x_{1},x_{2}\right) & = & \sum_{\left\{ i,j\right\} =1}^{2}\delta_{i}(x_{1})\delta_{j}(x_{2})H_{ij}(x_{1},x_{2}),
\end{eqnarray}
 with 
\begin{eqnarray}
H_{11}\left(x_{1},x_{2}\right) & = & b_{2}^{1}\left(x_{1},x\right)\sum_{i=1}^{3}\Omega_{i,i}\left(x\right)a_{1}^{i}\left(x,x_{1}\right)a_{2}^{i}\left(x,x_{2}\right)\nonumber \\
 & + & b_{3}^{6}\left(x_{1},x\right)\sum_{i=2}^{3}\Omega_{i,i}\left(x\right)a_{1}^{i}\left(x,x_{1}\right)a_{4}^{i}\left(x,x_{2}\right)\nonumber \\
 & + & \left[c_{6}^{1}\left(x_{1},x_{2}\right)+c_{8}^{1}\left(x_{1},x_{2}\right)\right]\sum_{i=1}^{3}\Omega_{i,i}\left(x\right)a_{6}^{i}\left(x,x_{1}\right)\nonumber \\
 & + & \left[c_{6}^{3}\left(x_{1},x_{2}\right)+c_{9}^{3}\left(x_{1},x_{2}\right)\right]\sum_{i=1}^{3}\Omega_{i,i}\left(x\right)a_{7}^{i}\left(x,x_{1}\right)\nonumber \\
 & - & b_{4}^{1}\left(x_{1},x_{2}\right)\sum_{i=1}^{3}\Omega_{i,i}\left(x\right)a_{2}^{3+i}\left(x,x_{1}\right),
\end{eqnarray}
\begin{eqnarray}
H_{12}\left(x_{1},x_{2}\right) & = & b_{2}^{1}\left(x_{1},x\right)\sum_{i=1}^{3}\Omega_{i,i}\left(x\right)a_{1}^{i}\left(x,x_{1}\right)a_{3}^{i}\left(x,x_{2}\right)\nonumber \\
 & + & b_{3}^{6}\left(x_{1},x\right)\sum_{i=2}^{3}\Omega_{i,i}\left(x\right)a_{1}^{i}\left(x,x_{1}\right)a_{5}^{i}\left(x,x_{2}\right)\nonumber \\
 & + & c_{9}^{1}\left(x_{1},x_{2}\right)\sum_{i=1}^{3}\Omega_{i,i}\left(x\right)a_{6}^{i}\left(x,x_{1}\right)\nonumber \\
 & + & c_{10}^{3}\left(x_{1},x_{2}\right)\sum_{i=1}^{3}\Omega_{i,i}\left(x\right)a_{7}^{i}\left(x,x_{1}\right)\nonumber \\
 & - & b_{5}^{1}\left(x_{1},x_{2}\right)\sum_{i=1}^{3}\Omega_{i,i}\left(x\right)a_{2}^{3+i}\left(x,x_{1}\right),
\end{eqnarray}
\begin{eqnarray}
H_{21}\left(x_{1},x_{2}\right) & = & b_{3}^{1}\left(x_{1},x\right)\sum_{i=1}^{3}\Omega_{i,i}\left(x\right)a_{1}^{i}\left(x,x_{1}\right)a_{2}^{i}\left(x,x_{2}\right)\nonumber \\
 & + & b_{4}^{6}\left(x_{1},x\right)\sum_{i=2}^{3}\Omega_{i,i}\left(x\right)a_{1}^{i}\left(x,x_{1}\right)a_{4}^{i}\left(x,x_{2}\right)\nonumber \\
 & + & \left[c_{7}^{1}\left(x_{1},x_{2}\right)+c_{10}^{1}\left(x_{1},x_{2}\right)\right]\sum_{i=1}^{3}\Omega_{i,i}\left(x\right)a_{6}^{i}\left(x,x_{1}\right)\nonumber \\
 & + & \left[c_{7}^{3}\left(x_{1},x_{2}\right)+c_{11}^{3}\left(x_{1},x_{2}\right)\right]\sum_{i=1}^{3}\Omega_{i,i}\left(x\right)a_{7}^{i}\left(x,x_{1}\right)\nonumber \\
 & - & b_{4}^{1}\left(x_{1},x_{2}\right)\sum_{i=1}^{3}\Omega_{i,i}\left(x\right)a_{3}^{3+i}\left(x,x_{1}\right),
\end{eqnarray}
and
\begin{eqnarray}
H_{22}\left(x_{1},x_{2}\right) & = & b_{3}^{1}\left(x_{1},x\right)\sum_{i=1}^{3}\Omega_{i,i}\left(x\right)a_{1}^{i}\left(x,x_{1}\right)a_{3}^{i}\left(x,x_{2}\right)\nonumber \\
 & + & b_{4}^{6}\left(x_{1},x\right)\sum_{i=2}^{3}\Omega_{i,i}\left(x\right)a_{1}^{i}\left(x,x_{1}\right)a_{5}^{i}\left(x,x_{2}\right)\nonumber \\
 & + & c_{11}^{1}\left(x_{1},x_{2}\right)\sum_{i=1}^{3}\Omega_{i,i}\left(x\right)a_{6}^{i}\left(x,x_{1}\right)\nonumber \\
 & + & c_{12}^{3}\left(x_{1},x_{2}\right)\sum_{i=1}^{3}\Omega_{i,i}\left(x\right)a_{7}^{i}\left(x,x_{1}\right)\nonumber \\
 & - & b_{5}^{1}\left(x_{1},x_{2}\right)\sum_{i=1}^{3}\Omega_{i,i}\left(x\right)a_{3}^{3+i}\left(x,x_{1}\right).
\end{eqnarray}

Now, in order to $\Psi_{2}\left(x_{1},x_{2}\right)$ given at (\ref{Psi2})
be an eigenstate of the transfer matrix (\ref{T}), all terms on (\ref{Action2})
but the first one must vanish. This is indeed true, provided that
the Bethe Ansatz equations of the 2-particle state, 
\begin{eqnarray}
\frac{\delta_{2}(x_{1})}{\delta_{1}(x_{1})} & = & -\frac{a_{1}^{1}(x_{1},x_{2})}{a_{1}^{2}(x_{1},x_{2})}\left(\frac{\Omega_{2,2}(x)a_{4}^{2}(x,x_{1})+\Omega_{3,3}(x)a_{4}^{3}(x,x_{1})}{\Omega_{2,2}(x)a_{5}^{2}(x,x_{1})+\Omega_{3,3}(x)a_{5}^{3}(x,x_{1})}\right),\label{BAE2a}\\
\frac{\delta_{2}(x_{2})}{\delta_{1}(x_{2})} & = & -\frac{a_{1}^{1}(x_{2},x_{1})}{a_{1}^{2}(x_{2},x_{1})}\left(\frac{\Omega_{2,2}(x)a_{4}^{2}(x,x_{2})+\Omega_{3,3}(x)a_{4}^{3}(x,x_{2})}{\Omega_{2,2}(x)a_{5}^{2}(x,x_{2})+\Omega_{3,3}(x)a_{5}^{3}(x,x_{2})}\right),\label{BAE2b}
\end{eqnarray}
are satisfied. Moreover, we can realize again that all dependence
of the Bethe Ansatz equations on the spectral parameter $x$ is only
apparent, as they should. Therefore, provided that the Bethe Ansatz
equations (\ref{BAE2a}) and (\ref{BAE2b}) are satisfied, $\Psi_{2}\left(x_{1},x_{2}\right)$
as given by (\ref{Psi2}) will be an eigenvector of the transfer matrix
(\ref{T}) with eigenvalue $\tau_{2}\left(x|x_{1},x_{2}\right)$ given
by (\ref{Eigen2}).

\subsection{The $n$-particle state }

From the previous cases we can figure out what should be the appropriated
$n$-particle state of the transfer matrix (\ref{T}). It follows
that $\Psi_{n}(x_{1},\ldots,x_{n})$ can be defined through a recurrence
relation of the form, 
\begin{eqnarray}
\Psi_{n}(x_{1},\ldots,x_{n}) & = & B_{1}(x_{1})\Psi_{n-1}\left(x_{1}^{\times}\right)\nonumber \\
 & + & \sum_{i=2}^{n}\lambda_{i}(x_{1},\ldots,x_{n})B_{2}(x_{1})\Psi_{n-2}\left(x_{1}^{\times},x_{i}^{\times}\right),\quad n\geq2,\label{PsiN}
\end{eqnarray}
where $\Psi_{0}$ is given by (\ref{Psi0}) and $\Psi_{1}(x_{1})$
is given by (\ref{Psi1}). We have also introduced the notation, 
\begin{equation}
\Psi_{n-1}\left(x_{i}^{\times}\right)=\prod_{k=1,k\neq i}^{n}B_{1}(x_{k}),\qquad\Psi_{n-2}\left(x_{i}^{\times},x_{j}^{\times}\right)=\prod_{k=1,k\neq\{i,j\}}^{n}B_{1}(x_{k}).
\end{equation}

The functions $\lambda_{i}(x_{1},\ldots,x_{n})$ appearing in (\ref{PsiN})
can be determined imposing the following exchange conditions,
\begin{eqnarray}
\Psi_{n}(x_{1},\ldots,x_{i},x_{i-1},\ldots,x_{n}) & = & \omega(x_{i-1},x_{i})\Psi_{n}(x_{1},\ldots,x_{i-1},x_{i},\ldots,x_{n}),\qquad2\leq i\leq n,
\end{eqnarray}
and using the commutations relations between the creator operators
in order to put them in a well-ordered form (see Appendix). This lead
us to the expressions, 
\begin{equation}
\omega(x_{i-1},x_{i})=b_{1}^{1}\left(x_{i+1},x_{i}\right),\qquad2\leq i\leq n,
\end{equation}
 and 
\begin{eqnarray}
\lambda_{k}(x_{1},\ldots,x_{n}) & = & -\prod_{j=2}^{k-1}\omega\left(x_{k},x_{j}\right)\left\{ b_{4}^{1}(x_{1},x_{k})\delta_{1}(x_{k})\prod_{i=2,i\neq k}^{n}a_{1}^{1}\left(x_{k},x_{i}\right)\right.\nonumber \\
 & + & \left.b_{5}^{1}(x_{1},x_{k})\delta_{2}(x_{k})\prod_{i=2,i\neq k}^{n}a_{1}^{2}\left(x_{k},x_{i}\right)\right\} ,\quad2\leq k\leq n.
\end{eqnarray}

The action of $t(x)$ on $\Psi_{n}(x_{1},\ldots,x_{n})$ can be computed
using many others commutation relations presented in the Appendix.
It follows that this action can be written as, 
\begin{eqnarray}
t(x)\Psi_{n}(x_{1},\ldots,x_{n}) & = & \tau_{n}\left(x|x_{1},\ldots,x_{n}\right)\Psi_{n}\left(x_{1},\ldots,x_{n}\right)\nonumber \\
 & + & \sum_{i=1}^{n}\beta_{i}^{1}\left(x_{1},\ldots,x_{n}\right)B_{1}(x)\Psi_{n-1}\left(x_{i}^{\times}\right)\nonumber \\
 & + & \sum_{i=1}^{n}\beta_{i}^{3}\left(x_{1},\ldots,x_{n}\right)B_{3}(x)\Psi_{n-1}\left(x_{i}^{\times}\right)\nonumber \\
 & + & \sum_{\{i,j\}=1,i<j}^{n}\beta_{i,j}^{2}\left(x_{1},\ldots,x_{n}\right)B_{2}(x)\Psi_{n-2}\left(x_{i}^{\times},x_{j}^{\times}\right),\label{ActionN}
\end{eqnarray}
where,
\begin{equation}
\tau_{n}\left(x|x_{1},\ldots,x_{n}\right)=\sum_{j=1}^{3}\Omega_{j,j}(x)\delta_{j}(x)\prod_{i=1}^{n}a_{1}^{j}(x,x_{i}),\label{EigenN}
\end{equation}
\begin{eqnarray}
\beta_{k}^{1}\left(x_{1},\ldots,x_{n}\right) & = & \prod_{l=1}^{k-1}\omega\left(x_{k},x_{l}\right)\delta_{1}\left(x_{k}\right)\sum_{j=1}^{3}\Omega_{j,j}\left(x\right)a_{2}^{j}\left(x,x_{k}\right)\prod_{i=1,i\neq k}^{n}a_{1}^{1}\left(x_{k},x_{i}\right)\nonumber \\
 & + & \prod_{l=1}^{k-1}\omega\left(x_{k},x_{l}\right)\delta_{2}\left(x_{2}\right)\sum_{j=1}^{3}\Omega_{j,j}\left(x\right)a_{3}^{j}\left(x,x_{i}\right)\prod_{i=1,i\neq k}^{n}a_{1}^{2}\left(x_{k},x_{i}\right),
\end{eqnarray}
\begin{eqnarray}
\beta_{k}^{3}\left(x_{1},\ldots,x_{n}\right) & = & \prod_{l=1}^{k-1}\omega\left(x_{k},x_{l}\right)\delta_{1}\left(x_{k}\right)\sum_{j=2}^{3}\Omega_{j,j}\left(x\right)a_{4}^{j}\left(x,x_{k}\right)\prod_{i=1,i\neq k}^{n}a_{1}^{1}\left(x_{k},x_{i}\right)\nonumber \\
 & + & \prod_{l=1}^{k-1}\omega\left(x_{k},x_{l}\right)\delta_{2}\left(x_{2}\right)\sum_{j=2}^{3}\Omega_{j,j}\left(x\right)a_{5}^{j}\left(x,x_{i}\right)\prod_{i=1,i\neq k}^{n}a_{1}^{2}\left(x_{k},x_{i}\right),
\end{eqnarray}
and 
\begin{eqnarray}
\beta_{i,j}^{2}\left(x_{1},\ldots,x_{n}\right) & = & \prod_{k=1}^{i-1}\omega\left(x_{k},x_{i}\right)\prod_{k=1}^{j-1}\omega\left(x_{k},x_{j}\right)\sum_{\left\{ p,q\right\} =1}^{2}\delta_{p}(x_{i})\delta_{q}(x_{j})H_{pq}\left(x_{i},x_{j}\right)\nonumber \\
 & \times & \prod_{k=1,k\neq\{i,j\}}^{n}a_{1}^{p}(x_{i},x_{k})\prod_{l=1,l\neq\{i,j\}}^{n}a_{1}^{q}(x_{i},x_{l}).
\end{eqnarray}

The requirement that $\Psi_{n}(x_{1},\ldots,x_{n})$ be an eigenstate
of the transfer matrix means that all terms in (\ref{ActionN}) that
are not proportional to $\Psi_{n}(x_{1},\ldots,x_{n})$ itself must
vanish. This lead us to the \emph{Bethe Ansatz equations} of the general
$n$-particle state:
\begin{equation}
\frac{\delta_{2}(x_{k})}{\delta_{1}(x_{k})}=-\left[\frac{\Omega_{2,2}(x)a_{4}^{2}(x,x_{k})+\Omega_{3,3}(x)a_{4}^{3}(x,x_{k})}{\Omega_{2,2}(x)a_{5}^{2}(x,x_{k})+\Omega_{3,3}(x)a_{5}^{3}(x,x_{k})}\right]\prod_{j=1,j\neq k}^{n}\frac{a_{1}^{1}\left(x_{k},x_{j}\right)}{a_{1}^{2}\left(x_{k},x_{j}\right)},\qquad1\leq k\leq n.
\end{equation}

\section{Explicit results}

Making use of the amplitudes of the $R$-matrix (\ref{R}), the expressions
of the elements of the diagonal reflection $K$-matrix (\ref{K})
and the coefficients of the commutation relations presented in the
appendix, we can explicitly write down the results of the boundary
\textsc{aba} for the Yang-Zhang model. It follows that the $n$-state
eigenvector of the transfer matrix (\ref{T}) is given by (\ref{PsiN})
where 
\begin{equation}
\omega\left(x,y\right)=\frac{\left(q^{2}x-y\right)}{\left(q^{2}y-x\right)}\frac{\left(qy+x\right)}{\left(qx+y\right)},
\end{equation}
 and 
\begin{eqnarray}
\lambda_{k}(x_{1},\ldots,x_{n}) & = & -\prod_{j=2}^{k-1}\frac{\left(q^{2}x_{k}-x_{j}\right)}{\left(q^{2}x_{j}-x_{k}\right)}\frac{\left(qx_{j}+x_{k}\right)}{\left(qx_{k}+x_{j}\right)}\times\nonumber \\
 &  & \left\{ \frac{\sqrt{q}\left(q^{2}-1\right)\left(x_{k}^{2}-1\right)x_{k}}{\left(qx_{k}+x_{1}\right)\left(q^{2}x_{k}^{2}-1\right)}\delta_{1}(x_{k})\prod_{i=2,i\neq k}^{n}\frac{\left(x_{i}x_{k}-1\right)\left(q^{2}x_{i}-x_{k}\right)}{\left(x_{i}-x_{k}\right)\left(q^{2}x_{i}x_{k}-1\right)}\right.\nonumber \\
 &  & \left.-\frac{1-q^{2}}{\sqrt{q}\left(qx_{1}x_{k}+1\right)}\delta_{2}(x_{k})\prod_{i=2,i\neq k}^{n}\frac{\left(q^{2}-1\right)\left(x_{i}^{2}-1\right)x_{i}}{\left(x_{i}-x_{k}\right)\left(q^{2}x_{i}^{2}-1\right)}\right\} ,\quad2\leq k\leq n.
\end{eqnarray}

The eigenvalues of the Sklyanin transfer matrix (\ref{T}) are given
by
\begin{eqnarray}
\tau_{n}\left(x|x_{1},\ldots,x_{n}\right) & = & \frac{\left(q^{3}x^{2}+1\right)\left[1-\frac{1}{2}(x-1)\left(\beta_{1,1}-\beta_{2,2}\right)\right]}{q^{3}x^{2}\left(qx^{2}+1\right)}\delta_{1}(x)\nonumber \\
 & \times & \left[\frac{\beta_{1,1}\left(q^{2}x^{2}-1\right)-2q^{2}x^{2}}{\left(q^{2}x-1\right)\left(\beta_{1,1}-\beta_{2,2}\right)-2}\right]\prod_{i=1}^{n}\left\{ \frac{\left(xx_{i}-1\right)\left(x-q^{2}x_{i}\right)}{\left(x-x_{i}\right)\left(q^{2}xx_{i}-1\right)}\right\} \nonumber \\
 & - & \frac{\left(qx-1\right)\left(qx+1\right)\left[1-\frac{1}{2}(x-1)\left(\beta_{1,1}-\beta_{2,2}\right)\right]}{q^{3}x^{3}(x^{2}-1)}\delta_{2}(x)\nonumber \\
 & \times & \left[\frac{\beta_{1,1}\left(q^{2}x^{2}-1\right)-2q^{2}x^{2}}{\left(q^{2}x-1\right)\left(\beta_{1,1}-\beta_{2,2}\right)-2}\right]\left[\frac{\left(qx+1\right)\left(\beta_{1,1}-\beta_{2,2}\right)+2qx}{\left(qx+1\right)\left(\beta_{1,1}-\beta_{2,2}\right)+2}\right]\nonumber \\
 & \times & \prod_{i=1}^{n}\left\{ -\frac{\left(xx_{i}-1\right)\left(x_{i}+qx\right)\left(q^{2}x_{i}-x\right)\left(q^{3}xx_{i}+1\right)}{q\left(x-x_{i}\right)\left(qx_{i}+x\right)\left(qxx_{i}+1\right)\left(q^{2}xx_{i}-1\right)}\right\} \nonumber \\
 & + & \left[\frac{\frac{1}{2}(x-1)\left(\beta_{1,1}-\beta_{2,2}\right)+x}{q^{2}x^{4}}\right]\delta_{3}(x)\nonumber \\
 & \times & \left[\frac{\beta_{1,1}\left(q^{2}x^{2}-1\right)-2q^{2}x^{2}}{\left(q^{2}x-1\right)\left(\beta_{1,1}-\beta_{2,2}\right)-2}\right]\left[\frac{\left(qx+1\right)\left(\beta_{1,1}-\beta_{2,2}\right)+2qx}{\left(qx+1\right)\left(\beta_{1,1}-\beta_{2,2}\right)+2}\right]\nonumber \\
 & \times & \prod_{i=1}^{n}\left\{ \frac{\left(x_{i}+qx\right)\left(q^{3}xx_{i}+1\right)}{q\left(qx_{i}+x\right)\left(qxx_{i}+1\right)}\right\} ,
\end{eqnarray}
where 
\begin{eqnarray}
\delta_{1}(x) & = & \left[1+\frac{1}{2}\left(x^{2}-1\right)\beta_{1,1}\right]\left(\frac{q^{2}x-1}{q^{2}/x-1}\right)^{N},\\
\delta_{2}(x) & = & -\frac{x\left(x^{2}-1\right)\left[1+\frac{1}{2}\left(x^{2}-1\right)\beta_{1,1}\right]}{\left(q^{2}x^{2}-1\right)}\left[\frac{\left(q^{2}x-1\right)\left(\beta_{1,1}-\beta_{2,2}\right)-2}{(x-1)\left(\beta_{1,1}-\beta_{2,2}\right)+2x}\right]\nonumber \\
 & \times & \frac{\left[q\left(x-1\right)\right]^{2N}}{\left[\left(q^{2}-x\right)\left(q^{2}x-1\right)/x\right]^{N}},\\
\delta_{3}(x) & = & -\frac{x^{2}\left(q+x^{2}\right)\left[1+\frac{1}{2}\left(x^{2}-1\right)\beta_{1,1}\right]}{q\left(qx^{2}+1\right)}\left[\frac{\left(q^{2}x-1\right)\left(\beta_{1,1}-\beta_{2,2}\right)-2}{\left(x-1\right)\left(\beta_{1,1}-\beta_{2,2}\right)+2x}\right]\nonumber \\
 & \times & \left[\frac{\left(qx+1\right)\left(\beta_{1,1}-\beta_{2,2}\right)+2}{\left(\beta_{1,1}-\beta_{2,2}\right)(q+x)+2x}\right]\frac{\left[q(x-1)(q+x)/\left(qx+1\right)\right]^{2N}}{\left[\left(q^{2}-x\right)\left(q^{2}x-1\right)/x\right]^{N}}.
\end{eqnarray}

Finally, the Bethe Ansatz equations are:
\begin{eqnarray}
\frac{\delta_{2}(x_{k})}{\delta_{1}(x_{k})} & = & \frac{qx_{k}\left(x_{k}^{2}-1\right)}{\left(q^{2}x_{k}^{2}-1\right)}\left[\frac{\left(qx_{k}+1\right)\left(\beta_{1,1}-\beta_{2,2}\right)+2}{\left(qx_{k}+1\right)\left(\beta_{1,1}-\beta_{2,2}\right)+2qx_{k}}\right]\nonumber \\
 & \times & \prod_{j=1,j\neq k}^{n}\left\{ \frac{q\left(qx_{j}+x_{k}\right)\left(qx_{j}x_{k}+1\right)}{\left(x_{j}+qx_{k}\right)\left(q^{3}x_{j}x_{k}+1\right)}\right\} ,\qquad1\leq k\leq n.
\end{eqnarray}

\section{Conclusion}

In this work we derived the boundary \textsc{aba} for the supersymetric
nineteen vertex model constructed from a three-dimensional boson free
representation $V$ of the twisted quantum affine Lie superalgebra
$U_{q}[\mathrm{osp}(2|2)^{(2)}]$. The $R$-matrix of this model was
introduced by Yang and Zhang in \cite{YangZhang1999} and the correponding
reflection $K$-matrices were derived by us recently in \cite{VieiraLima2017A}.
The eigenvalues and eigenvectors of Sklyanin's transfer matrix with
diagonal reflection $K$-matrices were determined, as well as the
corresponding Bethe Ansatz equations. Explicit results were also presented.

\ack{}{This work was supported in part by Brazilian Research Council (CNPq),
grant \#310625/2013-0 and FAPESP, grant \#2011/18729-1.}

\appendix

\section{The fundamental commutation relations}

To perform the boundary \textsc{aba}, we need to know how the diagonal
operators $D_{1}$, $D_{2}$ and $D_{3}$ pass through the creators
operators $B_{1}$, $B_{2}$ and $B_{3}$ (as an intermediate step,
we shall also need known how the $C$'s operators pass through the
$B$'s). These exchange rules are provided by the commutation relations
that can be derived from the so-called \emph{fundamental relation}
of the boundary \textsc{aba}: 
\begin{equation}
R_{12}(x/y)U_{1}(x)R_{21}(xy)U_{2}(y)=U_{2}(y)R_{12}(xy)U_{1}(x)R_{21}(x/y),\label{FR}
\end{equation}

In fact, writing $U(x)$ in the Lax representation as (\ref{U}),
and using the relations (\ref{F1}), (\ref{F2}), (\ref{F3}) and
(\ref{F4}), the following commutation relations ca be obtained (for
details about how these expressions are obtained, please see \cite{KurakLima2004,KurakLima2005}):
\begin{eqnarray}
D_{1}(x)B_{1}(y) & = & a_{1}^{1}(x,y)B_{1}(y)D_{1}(x)+a_{2}^{1}(x,y)B_{1}(x)D_{1}(y)+a_{3}^{1}(x,y)B_{1}(x)D_{2}(y)\nonumber \\
 & + & a_{6}^{1}(x,y)B_{2}(x)C_{1}(y)+a_{7}^{1}(x,y)B_{2}(x)C_{3}(y)+a_{8}^{1}(x,y)B_{2}(y)C_{1}(x),\label{D1B1}\\
D_{2}(x)B_{1}(y) & = & a_{1}^{2}(x,y)B_{1}(y)D_{2}(x)+a_{2}^{2}(x,y)B_{1}(x)D_{1}(y)+a_{3}^{2}(x,y)B_{1}(x)D_{2}(y)\nonumber \\
 & + & a_{4}^{2}(x,y)B_{3}(x)D_{1}(y)+a_{5}^{2}(x,y)B_{3}(x)D_{2}(y)+a_{6}^{2}(x,y)B_{2}(x)C_{1}(y)\nonumber \\
 & + & a_{7}^{2}(x,y)B_{2}(x)C_{3}(y)+a_{8}^{2}(x,y)B_{2}(y)C_{1}(x)+a_{9}^{2}(x,y)B_{2}(y)C_{3}(x),\label{D2B1}\\
D_{3}(x)B_{1}(y) & = & a_{1}^{3}(x,y)B_{1}(y)D_{3}(x)+a_{2}^{3}(x,y)B_{1}(x)D_{1}(y)+a_{3}^{3}(x,y)B_{1}(x)D_{2}(y)\nonumber \\
 & + & a_{4}^{3}(x,y)B_{3}(x)D_{1}(y)+a_{5}^{3}(x,y)B_{3}(x)D_{2}(y)+a_{6}^{3}(x,y)B_{2}(x)C_{1}(y)\nonumber \\
 & + & a_{7}^{3}(x,y)B_{2}(x)C_{3}(y)+a_{8}^{3}(x,y)B_{2}(y)C_{1}(x)+a_{9}^{3}(x,y)B_{2}(y)C_{3}(x),\label{D3B1}\\
D_{1}(x)B_{2}(y) & = & a_{1}^{4}(x,y)B_{2}(y)D_{1}(x)+a_{2}^{4}(x,y)B_{2}(x)D_{1}(y)+a_{3}^{4}(x,y)B_{2}(x)D_{2}(y)\nonumber \\
 & + & a_{4}^{4}(x,y)B_{2}(x)D_{3}(y)+a_{5}^{4}(x,y)B_{1}(x)B_{1}(y)+a_{6}^{4}(x,y)B_{1}(x)B_{3}(y),\label{D1B2}\\
D_{2}(x)B_{2}(y) & = & a_{1}^{5}(x,y)B_{2}(y)D_{2}(x)+a_{2}^{5}(x,y)B_{2}(x)D_{1}(y)+a_{3}^{5}(x,y)B_{2}(x)D_{2}(y)\nonumber \\
 & + & a_{4}^{5}(x,y)B_{2}(x)D_{3}(y)+a_{5}^{5}(x,y)B_{1}(x)B_{1}(y)+a_{6}^{5}(x,y)B_{1}(x)B_{3}(y)\nonumber \\
 & + & a_{7}^{5}(x,y)B_{3}(x)B_{1}(y)+a_{8}^{5}(x,y)B_{3}(x)B_{3}(y),\label{D2B2}\\
D_{3}(x)B_{2}(y) & = & a_{1}^{6}(x,y)B_{2}(y)D_{3}(x)+a_{2}^{6}(x,y)B_{2}(x)D_{1}(y)+a_{3}^{6}(x,y)B_{2}(x)D_{2}(y)\nonumber \\
 & + & a_{4}^{6}(x,y)B_{2}(x)D_{3}(y)+a_{5}^{6}(x,y)B_{1}(x)B_{1}(y)+a_{6}^{6}(x,y)B_{1}(x)B_{3}(y)\nonumber \\
 & + & a_{7}^{6}(x,y)B_{3}(x)B_{1}(y)+a_{8}^{6}(x,y)B_{3}(x)B_{3}(y),\label{D3B2}\\
D_{1}(x)B_{3}(y) & = & a_{1}^{7}(x,y)B_{3}(y)D_{1}(x)+a_{2}^{7}(x,y)B_{1}(y)D_{1}(x)+a_{3}^{7}(x,y)B_{1}(x)D_{1}(y)\nonumber \\
 & + & a_{4}^{7}(x,y)B_{1}(x)D_{2}(y)+a_{5}^{7}(x,y)B_{1}(x)D_{3}(y)+a_{6}^{7}(x,y)B_{2}(x)C_{1}(y)\nonumber \\
 & + & a_{7}^{7}(x,y)B_{2}(x)C_{3}(y)+a_{8}^{7}(x,y)B_{2}(y)C_{1}(x),\\
D_{2}(x)B_{3}(y) & = & a_{1}^{8}(x,y)B_{3}(y)D_{2}(x)+a_{2}^{8}(x,y)B_{1}(y)D_{2}(x)+a_{3}^{8}(x,y)B_{1}(x)D_{1}(y)\nonumber \\
 & + & a_{4}^{8}(x,y)B_{1}(x)D_{2}(y)+a_{5}^{8}(x,y)B_{1}(x)D_{3}(y)+a_{6}^{8}(x,y)B_{3}(x)D_{1}(y)\nonumber \\
 & + & a_{7}^{8}(x,y)B_{3}(x)D_{2}(y)+a_{8}^{8}(x,y)B_{3}(x)D_{3}(y)+a_{9}^{8}(x,y)B_{2}(x)C_{1}(y)\nonumber \\
 & + & a_{10}^{8}(x,y)B_{2}(x)C_{3}(y)+a_{11}^{8}(x,y)B_{2}(y)C_{1}(x)+a_{12}^{8}(x,y)B_{2}(y)C_{3}(x),\\
D_{3}(x)B_{3}(y) & = & a_{1}^{9}(x,y)B_{3}(y)D_{3}(x)+a_{2}^{9}(x,y)B_{1}(y)D_{3}(x)+a_{3}^{9}(x,y)B_{1}(x)D_{1}(y)\nonumber \\
 & + & a_{4}^{9}(x,y)B_{1}(x)D_{2}(y)+a_{5}^{9}(x,y)B_{1}(x)D_{3}(y)+a_{6}^{9}(x,y)B_{3}(x)D_{1}(y)\nonumber \\
 & + & a_{7}^{9}(x,y)B_{3}(x)D_{2}(y)+a_{8}^{9}(x,y)B_{3}(x)D_{3}(y)+a_{9}^{9}(x,y)B_{2}(x)C_{1}(y)\nonumber \\
 & + & a_{10}^{9}(x,y)B_{2}(x)C_{3}(y)+a_{11}^{9}(x,y)B_{2}(y)C_{1}(x)+a_{12}^{9}(x,y)B_{2}(y)C_{3}(x).
\end{eqnarray}

The commutation relation among the creators operators $B_{1}$, $B_{2}$
and $B_{3}$ themselves are: 
\begin{eqnarray}
B_{1}(x)B_{1}(y) & = & b_{1}^{1}(x,y)B_{1}(y)B_{1}(x)+b_{2}^{1}(x,y)B_{2}(y)D_{1}(x)+b_{3}^{1}(x,y)B_{2}(y)D_{2}(x)\nonumber \\
 & + & b_{4}^{1}(x,y)B_{2}(x)D_{1}(y)+b_{5}^{1}(x,y)B_{2}(x)D_{2}(y),\label{B1B1}\\
B_{2}(x)B_{1}(y) & = & b_{1}^{2}(x,y)B_{1}(y)B_{2}(x)+b_{2}^{2}(x,y)B_{2}(y)B_{1}(x)+b_{3}^{2}(x,y)B_{2}(y)B_{3}(x),\\
B_{3}(x)B_{1}(y) & = & b_{1}^{3}(x,y)B_{1}(y)B_{3}(x)+b_{2}^{3}(x,y)B_{1}(y)B_{1}(x)+b_{3}^{3}(x,y)B_{2}(y)D_{1}(x)\nonumber \\
 & + & b_{4}^{3}(x,y)B_{2}(y)D_{2}(x)+b_{5}^{3}(x,y)B_{2}(y)D_{3}(x)+b_{6}^{3}(x,y)B_{2}(x)D_{1}(y)\nonumber \\
 & + & b_{7}^{3}(x,y)B_{2}(x)D_{2}(y),\\
B_{1}(x)B_{2}(y) & = & b_{1}^{4}(x,y)B_{2}(y)B_{1}(x)+b_{2}^{4}(x,y)B_{2}(y)B_{3}(x)+b_{3}^{4}(x,y)B_{1}(y)B_{2}(x)\nonumber \\
 & + & b_{4}^{4}(x,y)B_{3}(y)B_{2}(x),\\
B_{2}(x)B_{2}(y) & = & B_{2}(y)B_{2}(x),\\
B_{3}(x)B_{2}(y) & = & b_{1}^{5}(x,y)B_{2}(y)B_{3}(x)+b_{2}^{5}(x,y)B_{1}(y)B_{2}(x)+b_{3}^{5}(x,y)B_{3}(y)B_{2}(x),\\
B_{1}(x)B_{3}(y) & = & b_{1}^{6}(x,y)B_{3}(y)B_{1}(x)+b_{2}^{6}(x,y)B_{1}(y)B_{1}(x)+b_{3}^{6}(x,y)B_{2}(y)D_{1}(x)\nonumber \\
 & + & b_{4}^{6}(x,y)B_{2}(y)D_{2}(x)+b_{5}^{6}(x,y)B_{2}(x)D_{1}(y)+b_{6}^{6}(x,y)B_{2}(x)D_{2}(y)\nonumber \\
 & + & b_{7}^{6}(x,y)B_{2}(x)D_{3}(y),\\
B_{2}(x)B_{3}(y) & = & b_{1}^{7}(x,y)B_{3}(y)B_{2}(x)+b_{2}^{7}(x,y)B_{1}(y)B_{2}(x)+b_{3}^{7}(x,y)B_{2}(y)B_{1}(x)\nonumber \\
 & + & b_{4}^{7}(x,y)B_{2}(y)B_{3}(x),\\
B_{3}(x)B_{3}(y) & = & b_{1}^{8}(x,y)B_{3}(y)B_{3}(x)+b_{2}^{8}(x,y)B_{1}(y)B_{1}(x)+b_{3}^{8}(x,y)B_{1}(y)B_{3}(x)\nonumber \\
 & + & b_{4}^{8}(x,y)B_{2}(y)D_{1}(x)+b_{5}^{8}(x,y)B_{2}(y)D_{2}(x)+b_{6}^{8}(x,y)B_{2}(y)D_{3}(x)\nonumber \\
 & + & b_{7}^{8}(x,y)B_{2}(x)D_{1}(y)+b_{8}^{8}(x,y)B_{2}(x)D_{2}(y)+b_{9}^{8}(x,y)B_{2}(x)D_{3}(y)\nonumber \\
 & + & b_{10}^{8}(x,y)B_{3}(y)B_{1}(x),
\end{eqnarray}
 Notice that we have chosen an appropriated order for the creators
operators, namely, that $B_{i}(x_{k})\prec B_{j}(x_{l})$ if, and
only if, $x_{k}<x_{l}$, for the indexes $\{i,j,k,l\}$ running from
$1$ to $3$. This order is important to the implementation of the
boundary \textsc{aba}, since we should use these commutation relations
until all operators be well ordered $-$ \emph{i.e.}, so that all
diagonal operators $D_{i}$, and all annihilator operators $C_{i}$,
be at right of the creator operators $B_{j}$ and, further, that among
the creator operators themselves, we always have $B_{i}\prec B_{j}$.

Finally, the commutation relations between the $C$'s and the $B$'s
are: 
\begin{eqnarray}
C_{1}(x)B_{1}(y) & = & c_{1}^{1}(x,y)B_{1}(y)C_{1}(x)+c_{2}^{1}(x,y)B_{1}(y)C_{3}(x)+c_{3}^{1}(x,y)B_{1}(x)C_{3}(y)\nonumber \\
 & + & c_{4}^{1}(x,y)B_{2}(x)C_{3}(y)+c_{5}^{1}(x,y)B_{2}(y)C_{2}(x)+c_{6}^{1}(x,y)D_{1}(y)D_{1}(x)\nonumber \\
 & + & c_{7}^{1}(x,y)D_{1}(y)D_{2}(x)+c_{8}^{1}(x,y)D_{1}(x)D_{1}(y)+c_{9}^{1}(x,y)D_{1}(x)D_{2}(y)\nonumber \\
 & + & c_{10}^{1}(x,y)D_{2}(x)D_{1}(y)+c_{11}^{1}(x,y)D_{2}(x)D_{2}(y),\label{C1B1}\\
C_{2}(x)B_{1}(y) & = & c_{1}^{2}(x,y)B_{3}(y)C_{2}(x)+c_{2}^{2}(x,y)C_{1}(y)D_{1}(x)+c_{3}^{2}(x,y)C_{1}(x)D_{1}(y)\nonumber \\
 & + & c_{4}^{2}(x,y)C_{1}(y)D_{2}(x)+c_{5}^{2}(x,y)C_{1}(x)D_{2}(y)+c_{6}^{2}(x,y)C_{1}(y)D_{3}(x)\nonumber \\
 & + & c_{7}^{2}(x,y)C_{3}(x)D_{1}(y)+c_{8}^{2}(x,y)C_{3}(x)D_{2}(y)+c_{9}^{2}(x,y)C_{1}(x)D_{1}(y)\nonumber \\
 & + & c_{10}^{2}(x,y)C_{3}(x)D_{1}(y)+c_{11}^{2}(x,y)C_{3}(y)D_{1}(x)+c_{12}^{2}(x,y)C_{1}(x)D_{2}(y)\nonumber \\
 & + & c_{13}^{2}(x,y)C_{3}(x)D_{2}(y)+c_{14}^{2}(x,y)C_{3}(y)D_{2}(x)+c_{15}^{2}(x,y)C_{3}(y)D_{3}(x),\label{C2B1}\\
C_{3}(x)B_{1}(y) & = & c_{1}^{3}(x,y)B_{1}(y)C_{3}(x)+c_{2}^{3}(x,y)B_{1}(y)C_{1}(x)+c_{3}^{3}(x,y)B_{1}(x)C_{3}(y)\nonumber \\
 & + & c_{4}^{3}(x,y)B_{2}(y)C_{2}(y)+c_{5}^{3}(x,y)B_{3}(x)C_{3}(y)+c_{6}^{3}(x,y)D_{1}(y)D_{1}(x)\nonumber \\
 & + & c_{7}^{3}(x,y)D_{1}(y)D_{2}(x)+c_{8}^{3}(x,y)D_{1}(y)D_{3}(x)+c_{9}^{3}(x,y)D_{1}(x)D_{1}(y)\nonumber \\
 & + & c_{10}^{3}(x,y)D_{1}(x)D_{2}(y)+c_{11}^{3}(x,y)D_{2}(x)D_{1}(y)+c_{12}^{3}(x,y)D_{2}(x)D_{2}(y)\nonumber \\
 & + & c_{13}^{3}(x,y)D_{3}(x)D_{1}(y)+c_{14}^{3}(x,y)D_{3}(x)D_{2}(y),\label{C3B1}\\
C_{1}(x)B_{2}(y) & = & c_{1}^{4}(x,y)B_{2}(y)C_{1}(x)+c_{2}^{4}(x,y)B_{2}(y)C_{3}(x)+c_{3}^{4}(x,y)B_{2}(x)C_{1}(y)\nonumber \\
 & + & c_{4}^{4}(x,y)B_{2}(x)C_{3}(y)+c_{5}^{4}(x,y)B_{1}(y)D_{1}(x)+c_{6}^{4}(x,y)B_{1}(y)D_{2}(x)\nonumber \\
 & + & c_{7}^{4}(x,y)B_{3}(y)D_{1}(x)+c_{8}^{4}(x,y)B_{3}(y)D_{2}(x)+c_{9}^{4}(x,y)B_{1}(x)D_{1}(y)\nonumber \\
 & + & c_{10}^{4}(x,y)B_{1}(x)D_{2}(y)+c_{11}^{4}(x,y)B_{1}(x)D_{3}(y)+c_{12}^{4}(x,y)B_{3}(x)D_{1}(y)\nonumber \\
 & + & c_{13}^{4}(x,y)B_{3}(x)D_{2}(y)+c_{14}^{4}(x,y)B_{3}(x)D_{3}(y),\\
C_{2}(x)B_{2}(y) & = & c_{1}^{5}(x,y)B_{2}(y)C_{2}(x)+c_{2}^{5}(x,y)B_{2}(x)C_{2}(y)+c_{3}^{5}(x,y)B_{1}(y)C_{1}(x)\nonumber \\
 & + & c_{4}^{5}(x,y)B_{1}(x)C_{1}(y)+c_{5}^{5}(x,y)B_{1}(y)C_{3}(x)+c_{6}^{5}(x,y)B_{1}(x)C_{3}(y)\nonumber \\
 & + & c_{7}^{5}(x,y)B_{3}(y)C_{1}(x)+c_{8}^{5}(x,y)B_{3}(y)C_{3}(x)+c_{9}^{5}(x,y)B_{3}(x)C_{3}(y)\nonumber \\
 & + & c_{10}^{5}(x,y)D_{1}(x)D_{2}(y)+c_{11}^{5}(x,y)D_{1}(x)D_{1}(y)+c_{12}^{5}(x,y)D_{1}(x)D_{1}(y)\nonumber \\
 & + & c_{13}^{5}(x,y)D_{2}(x)D_{1}(y)+c_{14}^{5}(x,y)D_{3}(x)D_{1}(y)+c_{15}^{5}(x,y)D_{1}(x)D_{3}(y)\nonumber \\
 & + & c_{16}^{5}(x,y)D_{1}(x)D_{2}(y)+c_{7}^{5}(x,y)D_{2}(x)D_{1}(y)+c_{18}^{5}(x,y)D_{2}(x)D_{2}(y)\nonumber \\
 & + & c_{19}^{5}(x,y)D_{2}(x)D_{2}(y)+c_{20}^{5}(x,y)D_{2}(x)D_{3}(y)+c_{21}^{5}(x,y)D_{1}(x)D_{3}(y)\nonumber \\
 & + & c_{22}^{5}(x,y)D_{3}(x)D_{1}(y)+c_{23}^{5}(x,y)D_{2}(x)D_{3}(y)+c_{24}^{5}(x,y)D_{3}(x)D_{2}(y)\nonumber \\
 & + & c_{25}^{5}(x,y)D_{3}(x)D_{3}(y),\\
C_{3}(x)B_{2}(y) & = & c_{1}^{6}(x,y)B_{2}(y)C_{3}(x)+c_{2}^{6}(x,y)B_{2}(x)C_{3}(y)+c_{3}^{6}(x,y)B_{1}(y)D_{1}(x)\nonumber \\
 & + & c_{4}^{6}(x,y)B_{1}(x)D_{1}(y)+c_{5}^{6}(x,y)B_{1}(y)D_{2}(x)+c_{6}^{6}(x,y)B_{1}(x)D_{2}(y)\nonumber \\
 & + & c_{7}^{6}(x,y)B_{1}(y)D_{3}(x)+c_{8}^{6}(x,y)B_{1}(x)D_{3}(y)+c_{9}^{6}(x,y)B_{2}(y)C_{1}(x)\nonumber \\
 & + & c_{10}^{6}(x,y)B_{2}(x)C_{1}(y)+c_{11}^{6}(x,y)B_{3}(y)D_{1}(x)+c_{12}^{6}(x,y)B_{3}(x)D_{1}(y)\nonumber \\
 & + & c_{13}^{6}(x,y)B_{3}(y)D_{2}(x)+c_{14}^{6}(x,y)B_{3}(x)D_{2}(y)+c_{15}^{6}(x,y)B_{3}(y)D_{3}(x)\nonumber \\
 & + & c_{16}^{6}(x,y)B_{3}(x)D_{3}(y),\\
C_{1}(x)B_{3}(y) & = & c_{1}^{7}(x,y)B_{3}(y)C_{1}(x)+c_{2}^{7}(x,y)B_{3}(x)C_{1}(y)+c_{3}^{7}(x,y)B_{1}(y)C_{1}(x)\nonumber \\
 & + & c_{4}^{7}(x,y)B_{1}(y)C_{3}(x)+c_{5}^{7}(x,y)B_{1}(x)C_{3}(y)+c_{6}^{7}(x,y)B_{2}(y)C_{2}(x)\nonumber \\
 & + & c_{7}^{7}(x,y)B_{3}(x)C_{3}(y)+c_{8}^{7}(x,y)D_{1}(x)D_{1}(y)+c_{9}^{7}(x,y)D_{1}(x)D_{1}(y)\nonumber \\
 & + & c_{10}^{7}(x,y)D_{2}(x)D_{1}(y)+c_{11}^{7}(x,y)D_{1}(x)D_{2}(y)+c_{12}^{7}(x,y)D_{1}(x)D_{3}(y)\nonumber \\
 & + & c_{13}^{7}(x,y)D_{1}(x)D_{2}(y)+c_{14}^{7}(x,y)D_{2}(x)D_{1}(y)+c_{15}^{7}(x,y)D_{2}(x)D_{2}(y)\nonumber \\
 & + & c_{16}^{7}(x,y)D_{2}(x)D_{3}(y)+c_{17}^{7}(x,y)D_{1}(x)D_{3}(y),\\
C_{2}(x)B_{3}(y) & = & c_{1}^{8}(x,y)B_{3}(y)C_{2}(x)+c_{2}^{8}(x,y)B_{3}(x)C_{2}(y)+c_{3}^{8}(x,y)C_{1}(y)D_{1}(x)\nonumber \\
 & + & c_{4}^{8}(x,y)C_{1}(x)D_{1}(y)+c_{5}^{8}(x,y)C_{1}(y)D_{2}(x)+c_{6}^{8}(x,y)C_{1}(x)D_{2}(y)\nonumber \\
 & + & c_{7}^{8}(x,y)C_{1}(y)D_{3}(x)+c_{8}^{8}(x,y)C_{1}(x)D_{3}(y)+c_{9}^{8}(x,y)C_{3}(y)D_{1}(x)\nonumber \\
 & + & c_{10}^{8}(x,y)C_{3}(x)D_{1}(y)+c_{11}^{8}(x,y)C_{3}(y)D_{2}(x)+c_{12}^{8}(x,y)C_{3}(x)D_{2}(y)\nonumber \\
 & + & c_{13}^{8}(x,y)C_{3}(x)D_{3}(y)+c_{14}^{8}(x,y)C_{1}(x)D_{1}(y)+c_{15}^{8}(x,y)C_{1}(y)D_{1}(x)\nonumber \\
 & + & c_{16}^{8}(x,y)C_{3}(x)D_{1}(y)+c_{17}^{8}(x,y)C_{3}(y)D_{1}(x)+c_{18}^{8}(x,y)C_{1}(x)D_{2}(y)\nonumber \\
 & + & c_{19}^{8}(x,y)C_{1}(y)D_{2}(x)+c_{20}^{8}(x,y)C_{3}(x)D_{2}(y)+c_{21}^{8}(x,y)C_{3}(y)D_{2}(x)\nonumber \\
 & + & c_{22}^{8}(x,y)C_{3}(x)D_{3}(y)+c_{23}^{8}(x,y)C_{3}(y)D_{3}(x),\\
C_{3}(x)B_{3}(y) & = & c_{1}^{9}(x,y)B_{3}(y)C_{3}(x)+c_{2}^{9}(x,y)B_{3}(x)C_{3}(y)+c_{3}^{9}(x,y)B_{1}(y)C_{1}(x)\nonumber \\
 & + & c_{4}^{9}(x,y)B_{1}(x)C_{1}(y)+c_{5}^{9}(x,y)B_{1}(y)C_{3}(x)+c_{6}^{9}(x,y)B_{1}(x)C_{3}(y)\nonumber \\
 & + & c_{7}^{9}(x,y)B_{2}(y)C_{2}(x)+c_{8}^{9}(x,y)B_{2}(x)C_{2}(y)+c_{9}^{9}(x,y)B_{3}(y)C_{1}(x)\nonumber \\
 & + & c_{10}^{9}(x,y)D_{1}(x)D_{1}(y)+c_{11}^{9}(x,y)D_{1}(x)D_{1}(y)+c_{12}^{9}(x,y)D_{2}(x)D_{1}(y)\nonumber \\
 & + & c_{13}^{9}(x,y)D_{1}(x)D_{2}(y)+c_{14}^{9}(x,y)D_{3}(x)D_{1}(y)+c_{15}^{9}(x,y)D_{1}(x)D_{3}(y)\nonumber \\
 & + & c_{16}^{9}(x,y)D_{1}(x)D_{2}(y)+c_{17}^{9}(x,y)D_{2}(x)D_{1}(y)+c_{18}^{9}(x,y)D_{2}(x)D_{2}(y)\nonumber \\
 & + & c_{19}^{9}(x,y)D_{2}(x)D_{2}(y)+c_{20}^{9}(x,y)D_{2}(x)D_{3}(y)+c_{21}^{9}(x,y)D_{1}(x)D_{3}(y)\nonumber \\
 & + & c_{22}^{9}(x,y)D_{3}(x)D_{1}(y)+c_{23}^{9}(x,y)D_{2}(x)D_{3}(y)+c_{24}^{9}(x,y)D_{3}(x)D_{2}(y)\nonumber \\
 & + & c_{25}^{9}(x,y)D_{3}(x)D_{3}(y).
\end{eqnarray}

The fundamental relation (\ref{FR}) also provides the commutation
relations between the operators $C_{i}$ and $D_{j}$, which can be
used to decrease a bit the number of terms of some commutation relations.
However, these supplementary commutation relations are not necessary
in the implementation of the boundary \textsc{aba} and they will not
be presented (the reader can consult references \cite{KurakLima2004,KurakLima2005}
for this purpose).

Next we will write down the coefficients of the commutation relations
above. We restrict ourselves only to the coefficients that appear
explicitly in the text; the others expressions (which are usually
very cumbersome and, as a mater of fact, not necessary) can also be
found following the lines of \cite{KurakLima2004,KurakLima2005}.
\begin{eqnarray}
a_{1}^{1}\left(x,y\right) & = & \frac{r_{1}(y/x)r_{2}(xy)}{r_{1}(xy)r_{2}(y/x)},\\
a_{2}^{1}\left(x,y\right) & = & -F_{1}(y)\frac{r_{5}(xy)}{r_{1}(xy)}-\frac{r_{2}(xy)s_{5}(y/x)}{r_{1}(xy)r_{2}(y/x)},\\
a_{3}^{1}\left(x,y\right) & = & -\frac{r_{5}(xy)}{r_{1}(xy)},\\
a_{6}^{1}\left(x,y\right) & = & \frac{r_{6}(xy)s_{5}(y/x)}{r_{1}(xy)r_{2}(y/x)},\\
a_{7}^{1}\left(x,y\right) & = & -\frac{r_{7}(xy)}{r_{1}(xy)},
\end{eqnarray}

\begin{eqnarray}
a_{1}^{2}\left(x,y\right) & = & \frac{\left[r_{5}(xy)s_{5}(xy)+r_{1}(xy)r_{4}(xy)\right]\left[r_{3}(x/y)r_{4}(x/y)-r_{6}(x/y)s_{6}(x/y)\right]}{r_{1}(xy)r_{2}(x/y)r_{2}(xy)r_{3}(x/y)},\\
a_{2}^{2}\left(x,y\right) & = & \left[\frac{s_{5}(x/y)s_{5}(xy)}{r_{2}(x/y)r_{2}(xy)}+F_{1}(x)\right]\left[\frac{F_{1}(y)r_{5}(xy)}{r_{1}(xy)}+\frac{r_{2}(xy)s_{5}(y/x)}{r_{1}(xy)r_{2}(y/x)}\right]\nonumber \\
 & - & \left[\frac{s_{5}(x/y)s_{6}(xy)}{r_{2}(xy)r_{3}(x/y)}-\frac{r_{1}(x/y)s_{5}(xy)s_{6}(x/y)}{r_{1}(xy)r_{2}(x/y)r_{3}(x/y)}\right]\frac{r_{6}(x/y)}{r_{2}(x/y)}\nonumber \\
 & + & \frac{F_{1}(y)r_{4}(xy)s_{5}(x/y)}{r_{2}(x/y)r_{2}(xy)}-\frac{r_{1}(x/y)r_{4}(x/y)s_{5}(xy)}{r_{1}(xy)r_{2}(x/y){}^{2}},\\
a_{3}^{2}\left(x,y\right) & = & \frac{r_{5}(xy)}{r_{1}(xy)}F_{1}(x)+\frac{s_{5}(x/y)\left[r_{5}(xy)s_{5}(xy)+r_{1}(xy)r_{4}(xy)\right]}{r_{1}(xy)r_{2}(x/y)r_{2}(xy)},\\
a_{4}^{2}\left(x,y\right) & = & \frac{r_{6}(xy)}{r_{2}(xy)}F_{1}(y)-\frac{r_{3}(xy)r_{6}(x/y)}{r_{3}(x/y)},\\
a_{5}^{2}\left(x,y\right) & = & \frac{r_{6}(xy)}{r_{2}(xy)},\\
a_{6}^{2}\left(x,y\right) & = & \frac{r_{1}(x/y)r_{4}(x/y)r_{6}(xy)s_{5}(xy)}{r_{1}(xy)r_{2}(x/y){}^{2}r_{2}(xy)}-\frac{F_{1}(x)r_{6}(xy)s_{5}(y/x)}{r_{1}(xy)r_{2}(y/x)},\\
a_{7}^{2}\left(x,y\right) & = & \frac{r_{7}(xy)}{r_{1}(xy)}F_{1}(x)+\frac{s_{5}(x/y)\left[r_{7}(xy)s_{5}(xy)-r_{1}(xy)r_{5}(xy)\right]}{r_{1}(xy)r_{2}(x/y)r_{2}(xy)},
\end{eqnarray}
 
\begin{eqnarray}
a_{1}^{3}\left(x,y\right) & = & \frac{r_{2}(x/y)\left(r_{6}(xy)s_{6}(xy)+r_{2}(xy){}^{2}\right)}{r_{2}(xy)r_{3}(x/y)r_{3}(xy)},\\
a_{2}^{3}\left(x,y\right) & = & \left[\frac{F_{1}(y)r_{5}(xy)}{r_{1}(xy)}+\frac{r_{2}(xy)s_{5}(y/x)}{r_{1}(xy)r_{2}(y/x)}\right]\left\{ \frac{s_{6}(x/y)s_{6}(xy)s_{5}(x/y)s_{5}(xy)}{r_{2}(x/y)r_{2}(xy)r_{3}(x/y)r_{3}(xy)}\right.\nonumber \\
 & + & \left.\frac{s_{7}(x/y)s_{7}(xy)}{r_{3}(x/y)r_{3}(xy)}-\left[F_{1}(x)-\frac{s_{5}(x/y)s_{5}(xy)}{r_{2}(x/y)r_{2}(xy)}\right]F_{3}(x)+F_{2}(x)\right\} \nonumber \\
 & + & \left[\frac{s_{5}(x/y)s_{6}(xy)}{r_{2}(xy)r_{3}(x/y)}-\frac{r_{1}(x/y)s_{5}(xy)s_{6}(x/y)}{r_{1}(xy)r_{2}(x/y)r_{3}(x/y)}\right]\left\{ F_{3}(x)\frac{r_{6}(x/y)}{r_{2}(x/y)}\right.\nonumber \\
 & - & \left.\frac{s_{6}(x/y)s_{6}(xy)r_{6}(x/y)}{r_{2}(x/y)r_{3}(x/y)r_{3}(xy)}-\frac{r_{2}(x/y)s_{6}(xy)}{r_{3}(x/y)r_{3}(xy)}\right\} \nonumber \\
 & + & \left[\frac{r_{4}(xy)s_{5}(x/y)s_{6}(x/y)s_{6}(xy)}{r_{2}(x/y)r_{2}(xy)r_{3}(x/y)r_{3}(xy)}-\frac{s_{5}(xy)s_{7}(x/y)}{r_{3}(x/y)r_{3}(xy)}\right]F_{1}(y)\nonumber \\
 & + & \frac{r_{2}(xy)s_{5}(x/y)s_{7}(xy)}{r_{1}(xy)r_{2}(x/y)r_{3}(x/y)r_{3}(xy)}-\frac{r_{1}(x/y)r_{4}(x/y)s_{5}(xy)s_{6}(x/y)s_{6}(xy)}{r_{1}(xy)r_{2}(x/y){}^{2}r_{3}(x/y)r_{3}(xy)},\\
a_{3}^{3}\left(x,y\right) & = & \frac{r_{4}(xy)s_{5}(x/y)s_{6}(x/y)s_{6}(xy)}{r_{2}(x/y)r_{2}(xy)r_{3}(x/y)r_{3}(xy)}-\frac{s_{5}(xy)s_{7}(x/y)}{r_{3}(x/y)r_{3}(xy)}\nonumber \\
 & - & \left[\frac{r_{5}(xy)s_{5}(xy)}{r_{1}(xy)}+r_{4}(xy)\right]\frac{s_{5}(x/y)F_{3}(x)}{r_{2}(x/y)r_{2}(xy)}\nonumber \\
 & + & \left[F_{2}(x)-F_{1}(x)F_{3}(x)\right]\frac{r_{5}(xy)}{r_{1}(xy)}\nonumber \\
 & + & \left[\frac{s_{5}(x/y)s_{5}(xy)s_{6}(x/y)s_{6}(xy)}{r_{2}(x/y)r_{2}(xy)r_{3}(x/y)r_{3}(xy)}+\frac{s_{7}(x/y)s_{7}(xy)}{r_{3}(x/y)r_{3}(xy)}\right]\frac{r_{5}(xy)}{r_{1}(xy)},\\
a_{4}^{3}\left(x,y\right) & = & \frac{r_{3}(x/y)s_{6}(x/y)\left[r_{6}(xy)s_{6}(xy)+r_{2}(xy){}^{2}\right]}{r_{2}(xy)r_{3}(x/y){}^{2}r_{3}(xy)}F_{1}(y)\nonumber \\
 & + & \frac{\left[r_{3}(xy)r_{6}(x/y)-r_{3}(x/y)r_{6}(xy)F_{1}(y)\right]}{r_{2}(xy)r_{3}(x/y)}F_{3}(x)\nonumber \\
 & - & \frac{s_{6}(xy)\left[r_{6}(x/y)s_{6}(x/y)+r_{2}(x/y){}^{2}\right]}{r_{2}(xy)r_{3}(x/y){}^{2}},\\
a_{5}^{3}\left(x,y\right) & = & \frac{s_{6}(x/y)\left[r_{6}(xy)s_{6}(xy)+r_{2}(xy){}^{2}\right]}{r_{2}(xy)r_{3}(x/y)r_{3}(xy)}-\frac{r_{6}(xy)}{r_{2}(xy)}F_{3}(x),\\
a_{6}^{3}\left(x,y\right) & = & \left[\frac{s_{6}(xy)s_{6}(x/y)}{r_{3}(x/y)r_{3}(xy)}-F_{3}(x)\right]\times\nonumber \\
 &  & \left\{ \left[\frac{r_{1}(x/y)r_{4}(x/y)s_{5}(xy)}{r_{2}(x/y)r_{2}(xy)s_{5}(y/x)}-F_{1}(x)\right]\frac{s_{5}(y/x)r_{6}(xy)}{r_{1}(xy)r_{2}(y/x)}\right.\nonumber \\
 & - & \left.\left[\frac{s_{5}(x/y)}{r_{3}(x/y)}+\frac{r_{1}(x/y)r_{6}(xy)s_{5}(xy)s_{6}(x/y)}{r_{1}(xy)r_{2}(x/y)r_{2}(xy)r_{3}(x/y)}\right]\left[\frac{r_{6}(x/y)}{r_{2}(x/y)}-\frac{r_{2}(x/y)s_{6}(xy)}{r_{3}(x/y)r_{3}(xy)}\right]\right\} \nonumber \\
 & - & \left[\frac{s_{5}(y/x)s_{5}(xy)}{r_{2}(y/x)r_{2}(xy)}+\frac{r_{1}(x/y)s_{7}(xy)}{r_{3}(x/y)r_{3}(xy)}\right]\frac{r_{6}(xy)s_{5}(x/y)}{r_{1}(xy)r_{2}(x/y)}+\frac{r_{6}(xy)s_{5}(y/x)}{r_{1}(xy)r_{2}(y/x)}F_{2}(x)\nonumber \\
 & + & \frac{\left[s_{6}(x/y)s_{6}(xy)F_{1}(x)-s_{7}(x/y)s_{7}(xy)\right]r_{6}(xy)s_{5}(y/x)}{r_{1}(xy)r_{2}(y/x)r_{3}(x/y)r_{3}(xy)},\\
a_{7}^{3}\left(x,y\right) & = & \left[\frac{s_{6}(x/y)s_{6}(xy)}{r_{3}(x/y)r_{3}(xy)}-F_{3}(x)\right]-\frac{r_{7}(xy)s_{6}(x/y)s_{6}(xy)}{r_{1}(xy)r_{3}(x/y)r_{3}(xy)}F_{1}(x)\nonumber \\
 & + & \frac{r_{7}(xy)}{r_{1}(xy)}F_{2}(x)+\left[\frac{s_{7}(xy)}{r_{1}(xy)}-r_{1}(xy)s_{7}(x/y)\right]\frac{r_{7}(xy)s_{7}(x/y)}{r_{3}(x/y)r_{3}(xy)},
\end{eqnarray}
 
\begin{eqnarray}
b_{1}^{1}(x,y) & = & \frac{s_{6}(x/y)\left[r_{6}(xy)s_{6}(xy)+r_{2}(xy){}^{2}\right]}{r_{2}(xy)r_{3}(x/y)r_{3}(xy)}-\frac{r_{6}(xy)}{r_{2}(xy)}F_{3}(x),\\
b_{2}^{1}(x,y) & = & \frac{r_{1}(y/x)\left[r_{3}(y/x)r_{6}(xy)F_{1}(x)-r_{3}(xy)r_{6}(y/x)\right]}{r_{2}(xy)\left[r_{3}(y/x)r_{4}(y/x)-r_{6}(y/x)s_{6}(y/x)\right]},\\
b_{3}^{1}(x,y) & = & \frac{r_{1}(y/x)r_{3}(y/x)r_{6}(xy)}{r_{2}(xy)\left[r_{3}(y/x)r_{4}(y/x)-r_{6}(y/x)s_{6}(y/x)\right]},\\
b_{4}^{1}(x,y) & = & \frac{r_{6}(xy)}{r_{2}(xy)}F_{1}(y)\nonumber \\
 & + & \frac{r_{3}(xy)\left[r_{6}(y/x)s_{7}(y/x)-r_{3}(y/x)s_{6}(y/x)\right]}{r_{2}(xy)\left[r_{3}(y/x)r_{4}(y/x)-r_{6}(y/x)s_{6}(y/x)\right]},\\
b_{5}^{1}(x,y) & = & \frac{r_{6}(xy)}{r_{2}(xy)},
\end{eqnarray}
\begin{eqnarray}
c_{6}^{1}(x,y) & = & \frac{r_{2}(xy)r_{5}(x/y)}{r_{1}(xy)r_{2}(x/y)}F_{1}(x)+\frac{s_{5}(xy)}{r_{1}(xy)},\\
c_{7}^{1}(x,y) & = & \frac{r_{2}(xy)r_{5}(x/y)}{r_{1}(xy)r_{2}(x/y)},\\
c_{8}^{1}(x,y) & = & -\frac{r_{2}(xy)s_{5}(x/y)}{r_{1}(xy)r_{2}(x/y)}F_{1}(y)-\frac{r_{5}(xy)}{r_{1}(xy)}F_{1}(x)F_{1}(y),\\
c_{9}^{1}(x,y) & = & -\frac{r_{5}(xy)}{r_{1}(xy)}F_{1}(x)-\frac{r_{2}(xy)s_{5}(x/y)}{r_{1}(xy)r_{2}(x/y)},\\
c_{10}^{1}(x,y) & = & -\frac{r_{5}(xy)}{r_{1}(xy)}F_{1}(y),\\
c_{11}^{1}(x,y) & = & -\frac{r_{5}(xy)}{r_{1}(xy)},
\end{eqnarray}
\begin{eqnarray}
c_{6}^{3}(x,y) & = & \frac{s_{5}(xy)s_{6}(x/y)}{r_{2}(xy)r_{3}(x/y)}\left[\frac{r_{2}(xy)r_{5}(x/y)}{r_{1}(xy)r_{2}(x/y)}F_{1}(x)+\frac{s_{5}(xy)}{r_{1}(xy)}\right]\nonumber \\
 & + & \frac{r_{4}(x/y)s_{6}(xy)}{r_{2}(xy)r_{3}(x/y)}F_{1}(x)-\frac{r_{3}(xy)r_{6}(x/y)}{r_{2}(xy)r_{3}(x/y)}F_{2}(x)-\frac{s_{6}(x/y)s_{7}(xy)}{r_{2}(xy)r_{3}(x/y)},\\
c_{7}^{3}(x,y) & = & -\frac{r_{3}(xy)r_{6}(x/y)}{r_{2}(xy)r_{3}(x/y)}F_{3}(x)+\frac{r_{5}(x/y)s_{5}(xy)s_{6}(x/y)}{r_{1}(xy)r_{2}(x/y)r_{3}(x/y)}+\frac{r_{4}(x/y)s_{6}(xy)}{r_{2}(xy)r_{3}(x/y)},\\
c_{8}^{3}(x,y) & = & -\frac{r_{3}(xy)r_{6}(x/y)}{r_{2}(xy)r_{3}(x/y)},\\
c_{9}^{3}(x,y) & = & -\frac{s_{5}(xy)s_{6}(x/y)}{r_{2}(xy)r_{3}(x/y)}\left[\frac{r_{2}(xy)s_{5}(x/y)}{r_{1}(xy)r_{2}(x/y)}F_{1}(y)+\frac{r_{5}(xy)}{r_{1}(xy)}F_{1}(x)F_{1}(y)\right]\nonumber \\
 & + & \frac{s_{6}(xy)s_{7}(x/y)}{r_{2}(xy)r_{3}(x/y)}F_{1}(y)+\frac{r_{6}(xy)}{r_{2}(xy)}F_{2}(x)F_{1}(y)-\frac{r_{4}(xy)s_{6}(x/y)}{r_{2}(xy)r_{3}(x/y)}F_{1}(x)F_{1}(y),\\
c_{10}^{3}(x,y) & = & -\frac{s_{5}(xy)s_{6}(x/y)}{r_{2}(xy)r_{3}(x/y)}\left[\frac{r_{5}(xy)}{r_{1}(xy)}F_{1}(x)+\frac{r_{2}(xy)s_{5}(x/y)}{r_{1}(xy)r_{2}(x/y)}\right]\nonumber \\
 & - & \frac{r_{4}(xy)s_{6}(x/y)}{r_{2}(xy)r_{3}(x/y)}F_{1}(x)+\frac{r_{6}(xy)}{r_{2}(xy)}F_{2}(x)+\frac{s_{6}(xy)s_{7}(x/y)}{r_{2}(xy)r_{3}(x/y)},\\
c_{11}^{3}(x,y) & = & -\frac{r_{4}(xy)s_{6}(x/y)}{r_{2}(xy)r_{3}(x/y)}F_{1}(y)-\frac{r_{5}(xy)s_{5}(xy)s_{6}(x/y)}{r_{1}(xy)r_{2}(xy)r_{3}(x/y)}F_{1}(y)+\frac{r_{6}(xy)}{r_{2}(xy)}F_{1}(y)F_{3}(x),\\
c_{12}^{3}(x,y) & = & \frac{r_{6}(xy)}{r_{2}(xy)}F_{3}(x)-\frac{r_{4}(xy)s_{6}(x/y)}{r_{2}(xy)r_{3}(x/y)}-\frac{r_{5}(xy)s_{5}(xy)s_{6}(x/y)}{r_{1}(xy)r_{2}(xy)r_{3}(x/y)}.
\end{eqnarray}

\section*{References}{}

\bibliographystyle{iopart-num}
\bibliography{references}

\end{document}